\documentclass[showpacs,nofootinbib,preprintnumbers,amsmath,amssymb]{revtex4}

\pdfoutput=1
\usepackage{dcolumn}
\usepackage{bm}
\usepackage{dsfont}
\usepackage{slashed}
\usepackage{graphicx}

\def\cyp{a}
\def\cyi{b}
\def\humboldt{c}
\def\jsc{d}
\def\wup{e}
\def\mit{f}

\newcommand{\Op}{\mathcal{O}}  
\newcommand{\eins}{\mathds{1}} 
\newcommand{\be}{\begin{equation}}
\newcommand{\ee}{\end{equation}}
\newcommand{\beq}{\begin{eqnarray}}
\newcommand{\eeq}{\end{eqnarray}}
\def\eqlab#1{\label{eq:#1}}

\def\Eqref#1{Eq.~(\ref{eq:#1})}

\begin{document}
\title{The electromagnetic form factors of the  $\Omega^{-}$ in lattice QCD}
\author{C. Alexandrou$^{(\cyp,\cyi)}$, T. Korzec$^{(\humboldt)}$, G. Koutsou$^{(\jsc,\wup)}$,  J. W. Negele$^{(\mit)}$,  Y. Proestos$^{(\cyi)}$}
\affiliation{
 {$^{(\cyp)}$ Department of Physics, University of Cyprus, P.O. Box 20537, 1678 Nicosia, Cyprus}\\
{$^{(\cyi)}$  Computation-based Science and Technology Research Center, The Cyprus Institute, P.O. Box 27456, 1645 Nicosia, Cyprus }\\
 {$^{(\humboldt)}$ Humboldt Universit\"at zu Berlin, Newtonstrasse 15, 12489 Berlin, Germany}\\
 {$^{(\jsc)}$ J\"ulich Supercomputing Center, Forschungszentrum J\"ulich, D-52425 J\"ulich, Germany}\\
 {$^{(\wup)}$ Bergische Universit\"at Wuppertal, Gaussstr. 20, D-42119 Wuppertal, Germany}\\
 {$^{(\mit)}$ Center for Theoretical Physics, 
Laboratory for
Nuclear Science and Department of Physics, Massachusetts Institute of
Technology, Cambridge, Massachusetts 02139, U.S.A.}\\
}
\begin{abstract}
We present results on the Omega baryon ($\Omega^{-}$)  electromagnetic form factors
using  $N_f=2+1$ domain-wall fermion configurations  
for three pion masses in the range of about 350 to 300 MeV.
 We compare results obtained 
 using domain wall fermions
 with those of a mixed-action (hybrid) approach, which combines
 domain wall valence quarks on staggered sea quarks,  for a
 pion mass of about 350 MeV. We pay particular attention
in the evaluation of the subdominant electric quadrupole form factor to sufficient 
accuracy
to exclude a zero value,
by constructing a sequential
source that isolates it from the dominant form factors.
The $\Omega^-$ magnetic moment, $\mu_{\Omega^{-}}$, the electric charge 
and magnetic 
 radius, $\langle r^{2}_{E0/M1} \rangle$, are extracted for these pion masses.
The electric quadrupole moment is determined for the first time using dynamical
quarks.
\end{abstract}
\pacs{11.15.Ha, 12.38.Gc, 12.38.Aw, 12.38.-t, 14.70.Dj}
\keywords{Lattice QCD, Hadron deformation, Form Factors, Omega Baryon}
\maketitle
\section{Introduction}
The structure of hadrons, such as size, shape and charge distribution 
can be probed by their electromagnetic form factors. 
The $\Omega^{-}$ baryon, consisting of three valence strange
quarks is, significantly more stable than other members of the baryon decuplet,
 such as the $\Delta$, with a life-time of the order of $10^{-10}$~s. 
This fact makes the calculation of its electromagnetic form factors  
particularly interesting since they are accessible to experimental 
measurements with smaller theoretical uncertainties. 
Its   magnetic dipole moment is measured to very good
accuracy, unlike those of the other decuplet baryons. A value of 
$\mu_{\Omega^{-}}=-2.02(5)$ 
is given in the PDG~\cite{PDG:2008}  in  units of nuclear magnetons ($\mu_N$).
 Within lattice QCD one can directly compute hadron form factors 
starting from the fundamental theory of the strong interactions. 
 Furthermore, higher order multipole moments, not detectable by current experimental setups, are accessible to lattice methods and can reveal important
information on the structure of the hadron. An example is
 the electric quadrupole moment, which detects deformation of a hadron state. 

In this work we calculate, for the first time, the electromagnetic form factors of 
the $\Omega^{-}$ baryon using dynamical 
domain-wall fermion configurations. For the calculation we use 
the fixed-sink approach, which enables the calculation of the form
 factors for all values and directions of the momentum 
transfer $\vec q$ concurrently. The main advantage of this approach 
is that it allows an increased statistical precision,
 while at the same time it provides  the full $Q^2$ dependence, where $Q^2=-q^2$. In order to obtain accurate results on the form factors we
construct optimized sources for the sequential inversion. 
 This is particularly important for the
subdominant electric quadrupole form factor, 
for which we construct  an appropriate source that isolates it
 from the two dominant form factors~\cite{Alexandrou:2008bn}.
This requires  extra  sequential inversions but it is essential in order to
determine 
the electric quadrupole form factor to good accuracy.

The form factors are  calculated using $N_f=2+1$ dynamical domain-wall fermion
  configurations at the three lowest pion masses
currently available, namely $m_\pi=350$ MeV,  $m_{\pi}=330$~MeV and $m_{\pi}=297$~MeV.
The results are compared to  those obtained with a hybrid action that 
uses domain wall valence quarks on staggered sea quarks simulated by the MILC 
collaboration~\cite{Orginos:1999cr}.

The paper is organized as follows: In Section II we provide the definitions
of the corresponding multipole form factors and describe the lattice setup to extract them. In 
Section III we discuss the results and in Section IV we give the conclusions.  
\section{Lattice techniques}
\subsection{Electromagnetic matrix element}
The $\Omega^-$ has spin and isospin 3/2 and therefore the decomposition
of the electromagnetic matrix element is the same as that of the $\Delta$. 
The on-shell $\Omega^{-}$ matrix element of 
the electromagnetic current $V^{\mu}$, is decomposed in terms of 
four independent Lorentz covariant vertex functions,
 $a_1(q^2)$, $a_2(q^2)$, $c_1(q^2)$ and $c_2(q^2)$, 
which  depend only on the squared momentum transfer $q^2=-Q^2=(p_{i}-p_{f})^2$.
 The initial and final  four-momentum 
are given by $p_{i}$ and $p_{f}$, 
respectively. 
In Minkowski spacetime these  covariant vertex functions are given  by~\cite{Nozawa:1990gt}
 \begin{align}\label{matrelem}
 \langle \Omega(p_f,s_f) | V^\mu | \Omega(p_i,s_i)\rangle =  \sqrt{\frac{m^{2}_{\Omega}}{E_\Omega(\vec{p_f})  E_\Omega(\vec{p_i})}}
&
\,\bar{u}_\sigma(p_f,s_f){\cal O}^{\sigma \mu \tau} u_\tau(p_i,s_i),\\[1mm]
\mathcal{O}^{\sigma \mu \tau} =-g^{\sigma \tau}\biggl[a_1(q^2) \gamma^\mu +\frac{a_2(q^2)}{2m_\Omega} \left(p_f^\mu + p_i^\mu\right)\biggr]
& -\frac{q^\sigma q^\tau}{4m_\Omega^2}\biggl[c_1(q^2)\gamma^\mu + \frac{c_2(q^2)}{2m_\Omega}\left(p_f^\mu+p_i^\mu\right)\biggr].
\end{align}
The rest mass and the energy of the particle are denoted by $m_\Omega$ and $E_\Omega,$ respectively. The initial and final spin-projections are given
 by $s_{i}$ and $s_{f}$, respectively. 
Recall also that  every vector component of the spin-$\frac{3}{2}$ 
Rarita-Schwinger vector-spinor $u_{\sigma}$ 
satisfies the Dirac equation,
$   \big(p_{\mu}\gamma^{\mu}-m_{\Omega}\big) u^\sigma(p,s) = 0$,   
along with the auxiliary conditions: $  \gamma_\sigma u^\sigma(p,s) = 0$ and $ p_\sigma u^\sigma(p,s)= 0$.
Additionally, the covariant vertex functions are linearly related 
to the (dimensionless) electric  $G_{E0}(q^2)$, $G_{E2}(q^2)$ and
  magnetic $G_{M1}(q^2)$, $G_{M3}(q^2)$ 
multipole form factors~\cite{Alexandrou:2008bn, Nozawa:1990gt}. 
Namely, the expressions relating the multipole form factors and the
 covariant vertex functions are given in Ref.~\cite{Nozawa:1990gt} 
and are quoted below for reference:
\begin{align}
\label{sachs1} G_{E0}&= (1+\frac{2}{3}\tau)[a_1+(1+\tau)a_2]-\frac{1}{3}\tau(1+\tau)[c_1+(1+\tau)c_2],\\
\label{sachs2} G_{E2}&=a_1+(1+\tau)a_2-\frac{1}{2}(1+\tau)[c_1+(1+\tau)c_2] ,\\
\label{sachs3} G_{M1}&=(1+\frac{4}{5}\tau)a_1-\frac{2}{5}\tau (1+\tau)c_1 ,\\
\label{sachs4} G_{M3}&= a_1-\frac{1}{2}(1+\tau)c_1,\\
\end{align}
where the positive quantity $\tau=-\frac{q^2}{4m_{\Omega}^{2}}.$
\subsection{Lattice setup}
We use gauge configurations generated by the RBC-UKQCD  collaborations
using $N_f=2+1$ domain-wall fermions~\cite{Allton:2008pn} and the 
Iwasaki gauge-action.   The simulations
are carried out on two lattices of
size $24^3\times 64$ at a pion mass of 330~MeV and  $32^3\times 64$  at
pion masses of 355~MeV and 297~MeV, respectively. The latter has a smaller lattice spacing and therefore
we will refer to it as the fine lattice.
For the $24^3\times 64$ lattice, 
or coarse lattice, the lattice spacing $a$, the light u- and d-quark mass as well as   the strange quark mass were
fixed by an iterative procedure using the $\Omega^-$, the pion and the kaon masses~\cite{Allton:2008pn} as inputs. 
The value obtained 
for the lattice spacing
is $a^{-1}=1.729(28)$~GeV~\cite{Allton:2008pn}. For the  fine lattice
the scale was fixed from the ratio of the pion decay constant,
 $f_\pi$ calculated
on the fine lattice to the one computed on the  $24^3\times 64$ 
at the same values
of the ratio $m_\pi/f_\pi$. The value found is
$a^{-1}=2.34(3)$~GeV~\cite{Syritsyn:2009mx}. 
 In addition, to these two lattices, we perform the calculation using a mixed-action
with  domain-wall valence quarks and staggered sea quarks.
The gauge configurations were produced by the MILC
 collaboration~\cite{Bernard:2001av,Aubin:2004ck}
using two degenerate flavors of light
staggered sea quarks 
and a  strange staggered sea quark fixed to its physical mass.
The lattice size is $28^3\times 64$ and the
 mass of the light quarks corresponds to a pion mass of 353~MeV. 
 The  lattice spacing is 0.124~fm as determined from the $\Upsilon^\prime-\Upsilon$ mass difference~\cite{Bernard:2001av}.
 For the valence quarks we use
domain wall fermions (DWF).
The valence strange-quark mass was set using the $N_F=3$ ensemble
by requiring the valence pseudoscalar mass to be equal to the mass 
of the Goldstone boson constructed using
 staggered quarks~\cite{WalkerLoud:2008bp}. 
Similarly the light quark valence mass is
tuned by adjusting the DWF pion mass to 
the taste-5 staggered Golstone boson pion.
The domain wall quark masses take the
values given in Table~\ref{Table:params}. Technical details
of this tuning procedure 
are given in Refs.~\cite{WalkerLoud:2008bp,Renner:2004ck,Hagler:2007xi}.

In all cases we used $N_5=16$,  which is 
what was used  in the simulation of the
 dynamical domain wall fermions.
We note that for the coarse lattice at the pion mass used here the
residual mass is large compared to the
bare quark mass and chiral symmetry breaking is expected.
The value of $N_5=16$ is also  used in the mixed action calculation
where  it was shown that
 the residual mass is 
 10\% of the bare quark mass, ensuring small chiral 
symmetry
breaking~\cite{Renner:2004ck}. 
In Table~\ref{Table:params} we provide details of the simulations, along with
the value of the  mass of the $\Omega^-$ obtained in this work 
as well as the value computed by other groups when available.

\begin{table}[ht]
\small
\begin{center}
\begin{tabular}{cccccccc}
\hline\hline 
$L_s^3\times L_T $& $N^{\mathrm{subd.}}_{\mathrm{confs}}$ & $N^{\mathrm{dom.}}_{\mathrm{confs}}$ & $a^{-1}$ [GeV] &  $m_{u,d}/m_s$ & $m_\pi$ [GeV] &  $m_N$  [GeV]& $m_\Omega$ [GeV]\\ 
\hline\hline
\multicolumn{7}{c}{$N_F=2+1$ domain wall fermions~\cite{Allton:2008pn}}\\
\hline
 $24^3\times 64$ & 200 & 200 & 1.729(28)&   0.005/0.04 & 0.329(1) &   1.154(7)~\cite{Aoki:2010xg} & 1.77(3) (1.758(9))~\cite{Allton:2008pn}\\
\hline
\multicolumn{7}{c}{$N_F=2+1$ domain wall fermions~\cite{Syritsyn:2009mx}}\\
\hline
$32^3\times 64$ & -- & 105 &  2.34(3) &   0.006/0.03 & 0.355(6) &  1.172(21) & 1.79(4) \\
$32^3\times 64$ & 200 & 120 & 2.34(3) &   0.004/0.03 & 0.297(5) &  1.109(21) & 1.76(2) \\
\hline
\multicolumn{7}{c}{Mixed action~\cite{Bratt:2010jn}} \\
\multicolumn{7}{c}{
DWF valence: $ am_{u,d} = 0.0138$, $am_s=0.081$}\\
\hline
$28^3\times 64$ &210 & 120 &  1.58(3) &  0.01/0.05 & 0.353(2) &1.191(19) &  1.78(3) (1.775(5))~\cite{WalkerLoud:2008bp}\\
\hline\hline
\end{tabular}
\end{center}
\caption{Parameters used
  in the calculation of the form factors. We give the 
 number of configurations $N^{\mathrm{subd.}}_{\mathrm{confs}}$ used to extract the subdominant electric quadrupole form factor
 $G_{E2}$, as well as the number of configurations used $N^{\mathrm{dom.}}_{\mathrm{confs}}$ to extract the dominant form factors for the various lattices employed in this study.
 The $\Omega^{-}$ hyperon mass
as determined in this work is given in the last column and it is compared with  the
value determined by the RBC-UKQCD collaboration and the LHPC for the
mixed action as given in parenthesis.}
\label{Table:params}
\end{table}
\subsection{Interpolating fields}
In order to calculate the on-shell  
matrix element
we utilize appropriate 
two- and three-point correlation functions.
An interpolating field operator with the quantum numbers of the $\Omega^-$
 baryon  is given by
\begin{align}\label{interpolator1}
\chi_{\sigma \alpha}(x)&= \epsilon^{abc}\, \mathbf{s}^{a}_{\alpha}\,\big(\mathbf{s}^{\!\mbox{\tiny T}b}_{\beta}\, [C\gamma_{\sigma}]_{\beta\gamma}\,  \mathbf{s}^{c}_{\gamma} \big),
\end{align}
where $C=\gamma_4 \gamma_2$ is the charge-conjugation matrix and $\sigma$ represents the vector index of the spin-$\frac{3}{2}$ spinor.
To ensure  ground state dominance at the shortest
possible Euclidean time separation
we perform a gauge invariant Gaussian smearing
on the strange quark fields that enter in the interpolating field,  
as described in Refs.~\cite{Alexandrou:1992ti,APEsmearing}:
\begin{align}
 {\mathbf s }_\beta(t,\vec x) = \sum_{ \vec y} [\mathds{1} + \alpha H(\vec x,\vec y; U)]^{n_{W}} \ s_\beta(t,\vec y),
\end{align}
\begin{align}
H(\vec x, \vec y; U)  = \sum_{\mu=1}^3 \left(U_\mu(\vec x,t)\delta_{\vec x, \vec y - \hat \mu} + U^{\dagger}_\mu(\vec x-\hat \mu, t) \delta_{\vec x,\vec y+\hat\mu} \right),
\end{align}
where $\mathbf{s}$  is the smeared s-quark field. 
The links $U_{\mu}(\vec x,\, t)$ entering the hopping matrix $H$ 
are APE-smeared gauge fields.
\begin{figure}[htb]
\begin{minipage}{8cm}
\hspace{-0.2cm}\includegraphics[width=\linewidth,height=\linewidth]{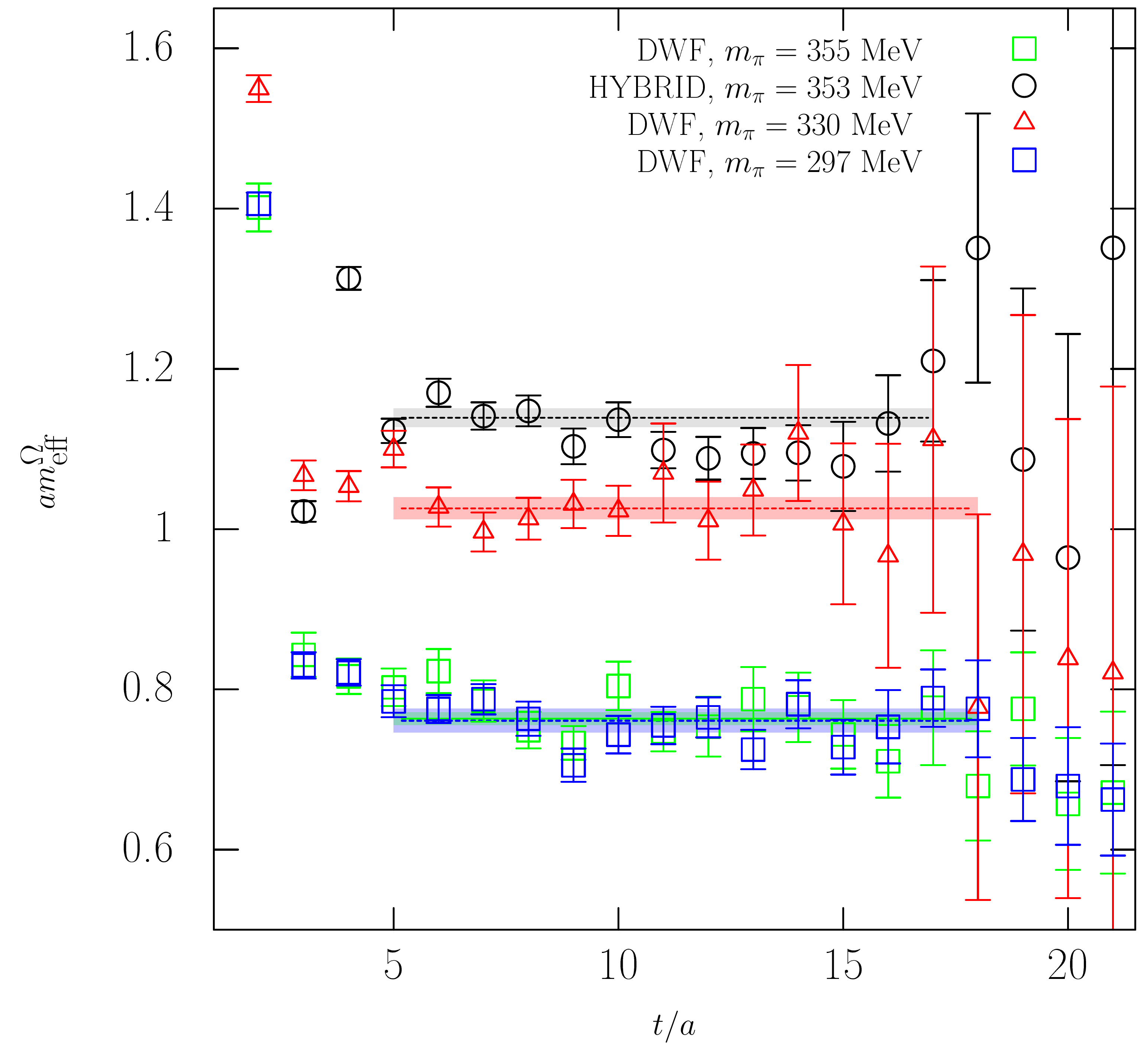}
\vspace*{-0.5cm}
\caption{The $\Omega^{-}$ effective mass and the fit to 
a constant plotted against the time separation
for each ensemble considered. The statistics used to extract the effective masses are summarized in Table~\ref{Table:params}.}
\label{meff}
\end{minipage}\hfill
\begin{minipage}{8cm}
\hspace{-0.2cm}\includegraphics[width=\linewidth,height=\linewidth]{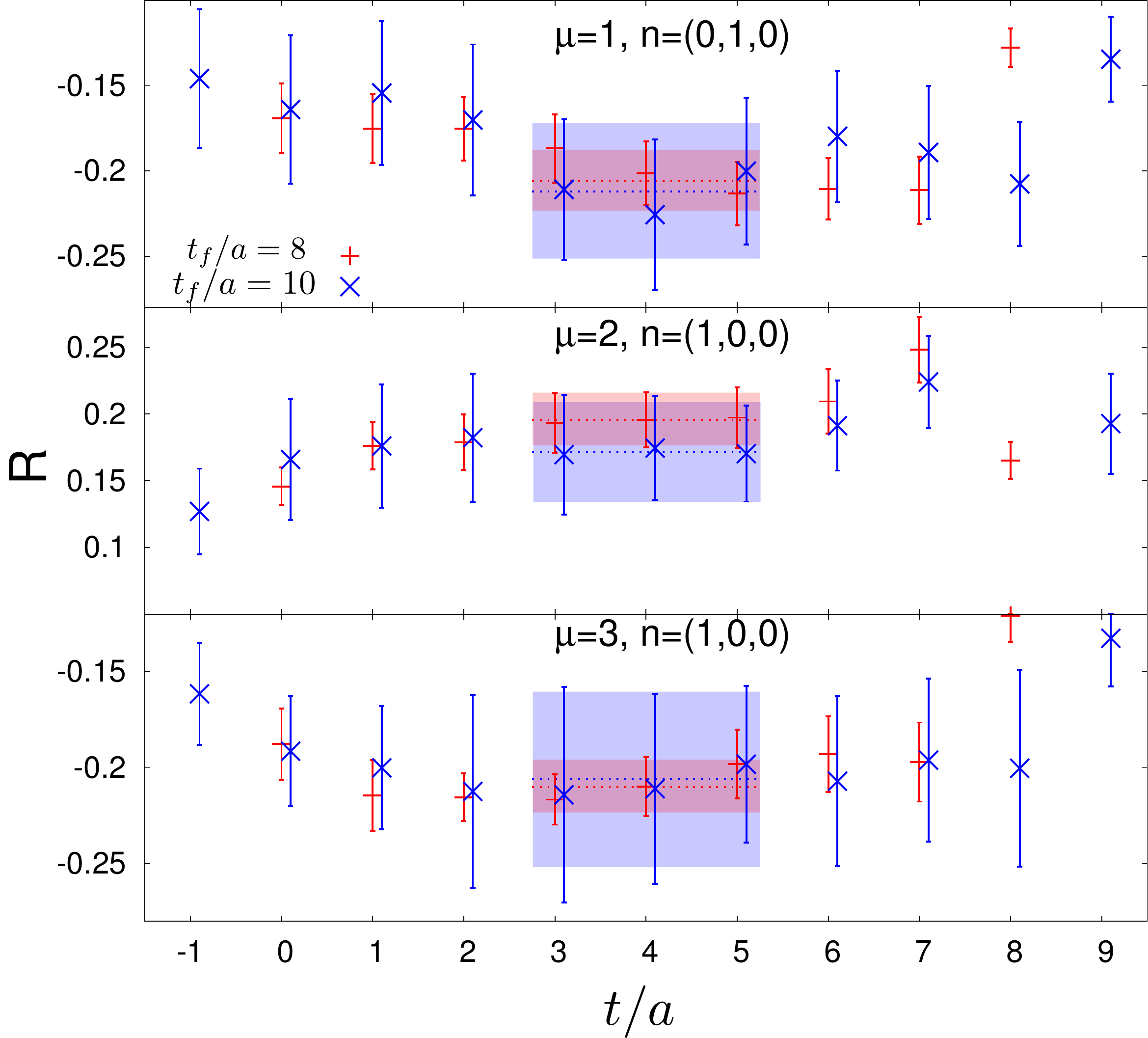}
\vspace*{0.5cm}\caption{
The ratio $R\equiv R_{\sigma \tau  \mu}(\Gamma,\vec q,t)$
 extracted for temporal source-sink separations $t_f/a=8$ and $t_f/a=10$, 
using 50 gauge configurations.   The results for $t_f/a=10$ 
are shifted to the left by one unit. 
We show results for current direction $\mu=1$ and $\mu=2,3$ 
and momenta 
 $\vec q$: $(0,1,0)\frac{2\pi}{L}$ and $(1,0,0)\frac{2\pi}{L}$, respectively. The bands correspond to the constant form fit  errors.
}
\label{figRRtype2}
\end{minipage}
\end{figure}
In particular, for DWF on  the coarse lattice 
 we have used the Gaussian smearing parameters $\alpha=5.026$ and $n_{W}=40$, 
while for the fine lattice the corresponding smearing parameters
 are  $\alpha=7.284$ and $n_{W}=84$. These are the same 
parameters as those  used  to ensure optimal filtering 
 of the nucleon state~\cite{Syritsyn:2009mx}.  

In Fig.~\ref{meff} we show
the results for the $\Omega^-$ effective mass calculated  from the two-point function ratio
$am_{eff}^{\Omega^{-}}(t)=-\log[G(t+1,{\vec{0}})/G(t,\vec{0})]$
  for the three different sets of configurations considered in this study,
The results are summarized in Table~\ref{Table:params}.

For the DWF simulations, on both the coarse and fine lattices  considered in this work, the resulting values
for the
$\Omega^-$ mass 
are  $1.77(3) \ \mathrm{GeV}$ and $1.76(2)\ \mathrm{GeV}$, respectively.
These values agree with the value found in Ref.~\cite{Allton:2008pn}.
The same agreement is obtained in the case of the hybrid action. 
In Ref.~\cite{Allton:2008pn} it was found that,  at the chiral limit,
 the $\Omega^-$ mass decreases by about 2\% its value at $am_{u,d}=0.005$.
Compared to the experimental value of   1.672~GeV~\cite{PDG:2008} the value
obtained at the physical point  is about 50~MeV higher 
 indicating that the strange quark mass 
is a few percent larger than 
the physical one
in these
simulations. 
\subsection{Two- and three-point Correlation functions}
The electromagnetic form factors can be extracted in lattice QCD  by constructing appropriate combinations of two- and three-point correlation functions.
 The corresponding lattice correlation functions are given by
\begin{eqnarray}
   G_{\sigma \tau}(\Gamma^\nu,\vec p, t) &=&\sum_{\vec x_f} e^{-i\vec x_f \cdot \vec p}\, 
   \Gamma^\nu_{\alpha'\alpha}\, \left\langle {\mathbf \chi}_{\sigma\alpha}(t,\vec x_f) \bar{\mathbf \chi}_{\tau\alpha'}(0, \vec 0) \right\rangle \eqlab{twopoint}\, , \\
   G_{\sigma\mu \tau}(\Gamma^\nu,\vec q, t) &=& \sum_{\vec x,\, \vec x_f} e^{i\vec x \cdot \vec q}\,
   \Gamma^\nu_{\alpha'\alpha}\, \left\langle {\mathbf \chi}_{\sigma\alpha}(t_f,\vec x_f) V_{\mu}(t,\vec x) \bar{\mathbf \chi}_{\tau\alpha'}(0, \vec 0)\right\rangle \, . \eqlab{threepointF}
\end{eqnarray}        
For our lattice  setup  we take a frame where
 the final $\Omega^-$-state is produced at rest i.e. $\vec p_f =\vec 0$.
Furthermore lattice calculations are carried out in a Euclidean space-time,
 and hence from here on  all expressions are given 
with Euclidean conventions~\cite{Montvay:1994cy}.
We use the local vector current $V_{\mu}$ carrying a momentum $\vec q=-\vec p_i,$ which is inserted at time $t$. The  
renormalization constant $Z_V$ is determined by the condition 
$G_E(0) = -1$. 
The $\Gamma$ matrices are given by
\begin{equation}
   \Gamma^4 = \frac{1}{4}(\eins + \gamma^4)\, , \qquad \qquad \Gamma^k =
   \frac{i}{4}(\eins + \gamma^4)\gamma_5\gamma_k\, , \qquad k=1, 2, 3\, .
\end{equation}
By inserting into the correlation functions a complete set of
energy momentum eigenstates  
\begin{equation}
   \sum_{n,p,\xi} \frac{M_n}{V\,E_{n(p)}} | n(p,\xi)\rangle \langle n(p,\xi) | =  \eins, \quad
   \eqlab{completeSet}
\end{equation}
with $\xi$ denoting
 all other quantum numbers, such as
spin,
one finds that the leading contributions for large Euclidean times $t$ 
and $t_f-t$ are
\begin{eqnarray}
   G_{\sigma \tau} (\Gamma^\nu,\vec p, t)         &=& \frac{M_\Omega}{E_{\Omega}(p)} \, |Z|^2\, e^{-E_{\Omega(p)}\, t}\,  
                                                      {\rm tr} \left[\Gamma^\nu \Lambda^E_{\sigma \tau}(p) \right] + {\rm excited\ states}\, ,\\
   G_{\sigma\mu\tau}(\Gamma^\nu,\vec q, t) &=&
                                                      \frac{M_\Omega}{E_{\Omega(p_i)}}\, |Z|^2 \, e^{-M_\Omega \,(t_f-t)}\,e^{-E_{\Omega}(p_i)\, t }\, 
{\rm tr}[\Gamma^\nu \Lambda^E_{\sigma \sigma'}(p_f) \Op^E_{\sigma' \mu \tau'} 
\Lambda^E_{\tau' \tau}(p_i)] \nonumber \\
&+& {\rm excited\ states} \, .
\end{eqnarray}
The leading time dependence and unknown overlaps of the $\Omega^-$
 state with the initial state
$\bar{J}_{\Omega}|0\rangle$ 
 in the three-point correlation function can be canceled out by 
forming appropriate ratios that involve  both the two-  and three-point
 functions. 
The ratio employed in this work is given by the following expression
\begin{align}\label{ratio1}
        R_{\sigma \mu \tau}(\Gamma,\vec q,t) &= \frac{G_{\sigma \mu \tau}(\Gamma^{\nu},\vec q,t)}{G_{k k}(\Gamma^4,\vec 0, t_f)}\ 
				         \sqrt{\frac{G_{kk}(\Gamma^4,\vec p_i, t_f-t)G_{kk}(\Gamma^4,\vec 0  ,t)G_{kk}(\Gamma^4,\vec 0,t_f)}
					            {G_{kk}(\Gamma^4,\vec 0, t_f-t)G_{kk}(\Gamma^4,\vec p_i,t)G_{kk}(\Gamma^4,\vec p_i,t_f)}}\, ,
\end{align}
where a summation over the repeated indices $k \,\,(k=1,2,3)$ is understood. This ratio becomes time-independent (displays a plateau) for large Euclidean time separations, that is
\begin{align}\label{ratio2}
   R_{\sigma \mu \tau}(\Gamma,\vec q,t) 
\stackrel{t_f-t \gg 1,t \gg 1}{\longrightarrow} \Pi_{\sigma\ \mu \tau}(\Gamma,\vec q) &= \mathcal{C}\
  \mathrm{Tr}\left[\Gamma\, \Lambda_{\sigma\sigma' }(p_f) \mathcal{O}_{{\sigma' }\mu{\tau' }} \Lambda_{\tau'  \tau}(p_i) \right] ,
\end{align}
\vspace*{-5mm}
\begin{align}
\label{prefactorratio}
\mathcal{C}&= \sqrt{\frac{3}{2}}\left[\frac{2 E_{\Omega}(\vec q)}{m_\Omega} 
                          +\frac{2 E^2_{\Omega}(\vec q)}{m^2_\Omega} 
                          +\frac{  E^3_{\Omega}(\vec q)}{m^3_\Omega} 
                          +\frac{  E^4_{\Omega}(\vec q)}{m^4_\Omega} \right]^{-\frac{1}{2}}. 
\end{align} 
It is understood that the trace acts in spinor-space, while the Rarita-Schwinger spin sum, expressed in Euclidean space,  is given by
\begin{align}\label{RSeuclidean}
   \Lambda_{\sigma\tau}(p)&\equiv \sum_{s} u_\sigma(p,s) \bar{u}_\tau(p,s) = - \frac{-i\slashed{p}+m_\Omega}{2m_\Omega}\left[
   \delta_{\sigma\tau}-\frac{\gamma_\sigma\gamma_{\tau}}{3}
                               +\frac{2p_\sigma p_{\tau}}{3m_\Omega^2} 
- i \frac{p_\sigma\gamma_{\tau}-p_{\tau}\gamma_\sigma}{3m_\Omega} \right] \, .
\end{align}
The electromagnetic form factors are extracted by fitting
$  R_{\sigma \tau \mu}(\Gamma,\vec q,t)$
 in the plateau region determined by $ \Pi_{\sigma\ \tau}^{\ \mu}(\Gamma,\vec q)$.

Since we are evaluating the correlator of \Eqref{threepointF} using  
sequential inversions through the sink~\cite{Dolgov:2002zm}, 
a separate set of inversions is necessary for every choice of vector and Dirac-indices.
The total of $256$ combinations arising from the vector indices of the $\Omega^-$ and the choice of $\Gamma$ matrices, as can be inferred from \Eqref{threepointF},  is beyond our computational resources, and hence we concentrate on a few 
carefully chosen combinations given below:
\begin{eqnarray}
   \Pi_\mu^{(1)}(\vec q) &=& \sum \limits_{j,k,l=1}^3 \epsilon_{jkl}\Pi_{j\mu
   k}(\Gamma^4, \vec q) \eqlab{comb1}  \\
   &=& G_{M1}\
   \frac{5i(E_{\Omega}+M_\Omega)\mathcal{C}}{18M_\Omega^2}\left[\delta_{1,\mu}(q_3-q_2)
   + \delta_{2,\mu}(q_1-q_3) + \delta_{3,\mu}(q_2-q_1)\right], \nonumber
\end{eqnarray}
\begin{eqnarray}
 \Pi_\mu^{(2)}(\vec q) &=& \sum \limits_{k=1}^3 \Pi_{k\mu k}(\Gamma^4, \vec
   q) \eqlab{comb2} \\
   &=&  -G_{E0}\ \frac{(E_{\Omega}+2M_\Omega)\mathcal{C}}{3M_\Omega^2}   \left[(M_\Omega+E_\Omega) \delta_{4,\mu}+iq_\mu(1-\delta_{4,\mu}) \right]\nonumber \\
   &-& G_{E2}\
   \frac{(E_{\Omega}-M_\Omega)^2\mathcal{C}}{9M_\Omega^3}\left[(M_\Omega+E_\Omega)
   \delta_{4,\mu}+iq_\mu(1-\delta_{4,\mu}) \right], \nonumber
\end{eqnarray}
\begin{eqnarray}
   \Pi_\mu^{(3)}(\vec q) &=& \sum \limits_{j,k,l=1}^3 \epsilon_{jkl}\Pi_{j\mu
   k}(\Gamma^j, \vec q) \eqlab{comb3} \\
   &=& G_{E2}\ \frac{-i\mathcal{C}}{3M_\Omega^2(E_{\Omega}+M_\Omega)} (q_1q_2 + q_2q_3 +
   q_3q_1) \nonumber \\
&&\hspace{3.5cm} \times
\left[(M_\Omega+E_\Omega)\delta_{4,\mu}+iq_\mu(1-\delta_{4,\mu}) \right] \nonumber \\
   &+& G_{M1}\ \frac{\mathcal{C}}{6M_\Omega^2(E_\Omega+M_\Omega)} \sum_{k=1}^3 
            \delta_{k,\mu}\, q_1 q_2 q_3 \left(2-\frac{q_1+q_2+q_3-q_k}{q_k} \right) \nonumber \\
   &+& G_{M3}\ \frac{\mathcal{C}}{30M_\Omega^3(E_\Omega+M_\Omega)}\sum_{k=1}^3
            \delta_{k,\mu}\biggl[(16E_\Omega+14M_\Omega)q_1 q_2 q_3 \nonumber \\
   &&\hspace{3.5cm} - 10M_\Omega(q_1 q_2 + q_2 q_3 + q_3 q_1) q_k 
\nonumber \\
   &&\hspace{3.5cm} -(8E_\Omega+7M_\Omega)\frac{q_1 q_2
   q_3}{q_k}(q_1+q_2+q_3-q_k)\biggr] \nonumber \, ,
\end{eqnarray}
where the kinematical factor ${\cal C}$  is given in~Eq.~(\ref{prefactorratio}).
As expected, current conservation $q_\mu \Pi_\mu = 0$ is manifest in the right hand side
of the equations. 
From these expressions all the multipole form factors can be extracted. For instance~\Eqref{comb1} 
is proportional to $G_{M1}$, while~\Eqref{comb3} isolates $G_{E2}$ for $\mu=4$.
Furthermore, these combinations are optimal in the sense that
all momentum directions, each of which is statistically different, contributes
to a given $Q^2$-value. This symmetric construction yields a 
better estimator for the $\Omega^-$-matrix elements than methods 
where only one momentum-vector is accessible.

In this paper, we consider only connected contributions to the three-point function. These are calculated by
performing sequential inversions through the sink, which necessitates
fixing the quantum numbers of the initial and final states as well as the
time separation between the source and the sink.
The optimal combinations given in \Eqref{comb1} - \Eqref{comb3},
from which $G_{E0}$, $G_{M1}$ and $G_{E2}$ are determined, can be implemented
by an appropriate sink construction which
requires only one sequential inversion for each of the three types of
combinations. No optimal
sink is considered for the octupole magnetic form factor 
in this work. Although it can and has been extracted, the results exhibit large errors
and are consistent with zero. We therefore refrain from presenting this specific
form factor.
The matrix element for all the different directions of $\vec q$ 
and for all four directions $\mu$ of the current can then be computed yielding 
an over-constrained system of linear equations which can be solved for the form factors in the least squares sense. 
A singular value decomposition of the coefficient matrix is utilized 
to find the least squares solution.
The statistical errors are found by a jack-knife procedure, which takes care
of any possible autocorrelations between gauge configurations.

As  already mentioned,
 the three-point function of the connected part  is calculated by
 performing sequential inversions through the sink. This requires 
fixing the temporal source-sink separation. In order, to determine
the smallest time separation that is still sufficiently large to damp
excited state contributions, we perform the calculation at two values of
the sink-source separation. We use $t_f/a=8$ and $t_f/a=10$
 for the DWF configurations corresponding to the coarse  lattice
 spacing $a=0.114$~fm. We  compare in Fig.~\ref{figRRtype2}
the  results for the plateaus   $\Pi_{\sigma \tau \mu}(\Gamma,\vec q)$,
 for a few selected directions of the current and for low momentum $\vec q$
 values for these two sink-source time separations. As can be seen, the
plateau values at $t_f/a=10$ are consistent with the smaller time separation
the latter exhibiting about half the statistical error. We therefore use $t_f/a=8$ or
$t_f=0.91$~fm as source sink separation.  For the fine DWF lattice 
 the inversions were performed for $t_f/a=12,$ which corresponds 
to about  $t_f=1.008$~fm. Similarly for the hybrid scheme 
the time separation was taken to be at $t_f/a=8$ or $t_f=0.992$~fm.
\section{Results}
We use the local electromagnetic current,  
$V^{\mu}=-\frac{1}{3}\bar{s}\gamma^{\mu}s$,
which requires a renormalization factor $Z_{V}$ 
to be included. The vector current renormalization constant is determined from the lattice calculation by the requirement that
\begin{align}
Z_{V}G_{E0}(0)=-1,
\label{renormZV}
\end{align}
where -1 is the charge of $\Omega^{-}$.
The values of $Z_V$ extracted using Eq.~(\ref{renormZV}) are given
in Table~\ref{ResultsTable}, where the errors shown are statistical. 
For the coarse lattice with DWF, 
the  value of $Z_{V}=0.7161(1)$ is calculated~\cite{Aoki:2007xm} 
from the pion decay constant. For the
 fine lattice $Z_V$ was fixed using the nucleon electric form factor~\cite{Syritsyn:2009mx} with values $Z_{V}=0.7468(39)$ at $m_{\pi}=297$~MeV  and $Z_{V}=0.7479(22)$ at $m_{\pi}=355$~MeV.
For the mixed-action~\cite{Bratt:2010jn} with $m_{\pi}=353$ MeV 
the value of the current renormalization constant $Z_{V}=1.1169$ is obtained 
by dividing the unrenormalized isovector current with  the forward matrix element. These values differ by about 1\%-2\% 
with the ones found using Eq.~(\ref{renormZV}). This discrepancy indicates
systematic errors on the 2\% level.
\subsection{Electric charge form factor}
\begin{figure}[h]
\begin{center}
\hspace{-0.2cm}\includegraphics[width=0.6\linewidth,height=0.5\linewidth]{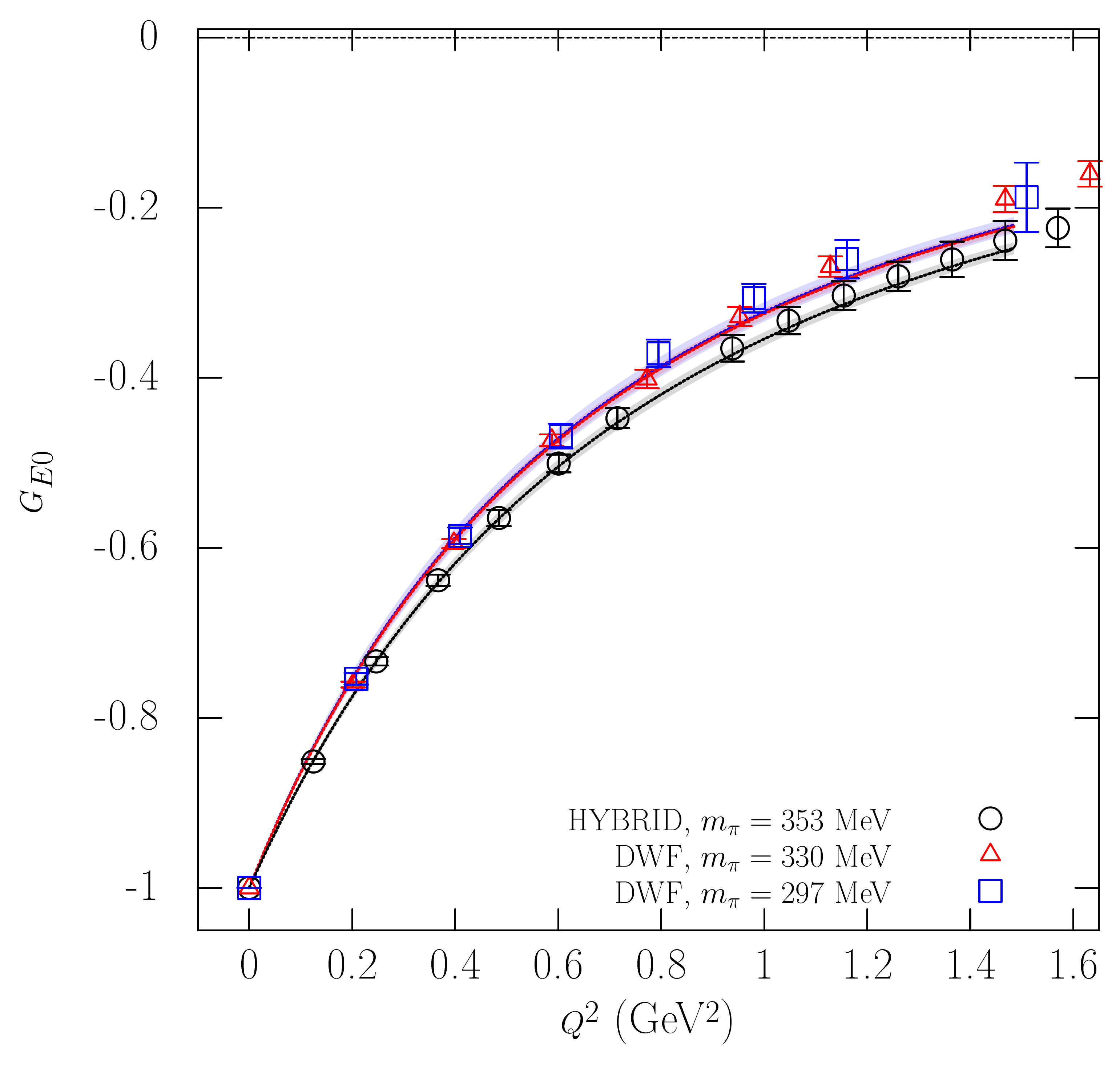} 
\vspace*{-0.5cm}
   \caption{The electric charge form factor $G_{E0}(Q^2)$ computed at  $m_{\pi}=330\ \mathrm{MeV}$ and at $m_{\pi}=297\ \mathrm{MeV}$. The lines describe the \emph{dipole fits} given by Eq.~(\protect \ref{ge0dipoleform}), while the bands show the corresponding errors to the fits.}\label{figge0HYBRBCRBC}
\end{center}
\end{figure}

Our results for the electric charge form factor, $G_{E0}(Q^2)$,
 are depicted in Fig.~\ref{figge0HYBRBCRBC} for the fine and coarse lattice using DWF and for the mixed action.
Results using the mixed action have consistently smaller values.
This
can be attributed either to cut-off effects or to a small dependence on the mass of the light sea quack mass.
In order to  check, we perform a calculation using DWF at $m_\pi=355$~MeV on the fine lattice for the
magnetic dipole form factor. This will be discussed in next section.
In Fig.~\ref{figge0HYBRBCRBC} we show fits to a dipole.  As can be seen, the momentum dependence of this form factor is adequately described in all cases  by a one-parameter dipole form
\begin{align}\label{ge0dipoleform}
   G_{E0}(Q^2) &= -\frac{1}{\big(1+ \frac{Q^2}{\Lambda_{E0}^{2}}\big)^2}.
\end{align}  
In the non-relativistic limit the slope of the above dipole form evaluated 
at momentum transfer $Q^2=0$, is related to the electric charge 
mean square radius by
\begin{align}\label{rmrradius}
   \left\langle r^{2}_{E0} \right\rangle = -\frac{6}{G_{E0}(0)} \frac{d}{dQ^2} G_{E0}(Q^2)\bigg|_{Q^2=0} \, .
\end{align}

From the dipole fit to the coarse DWF lattice data we determine $\Lambda_{E0}$ and
obtain
 a value of $\langle r^{2}_{E0}\rangle=0.353(8)\ \mathrm{fm}^2$, while for the fine DWF lattice the corresponding value turns out to be $\langle r^{2}_{E0}\rangle=0.355(14)\ \mathrm{fm}^2$~\footnote{Note the different sign as compared to 
Ref.~\cite{Boinepalli:2009sq} since we here divide by $G_E(0)=-1$.}.
These values  are slightly greater in magnitude than the one 
reported in Ref.~\cite{Boinepalli:2009sq}, which was obtained in
 a quenched lattice QCD calculation. 
 The discrepancy may originate from unquenching effects or 
 pronounced light quark mass dependence since the pion mass
used in the quenched study of  Ref.~\cite{Boinepalli:2009sq} is larger than 
what used here.
The results
for the  $ \left\langle r^{2}_{E0} \right\rangle$ are given in Table~\ref{ResultsTable}. 
\subsection{Magnetic dipole form factor}
%

\begin{figure}[h]
\begin{minipage}{8cm}
\includegraphics[width=\linewidth,height=\linewidth]{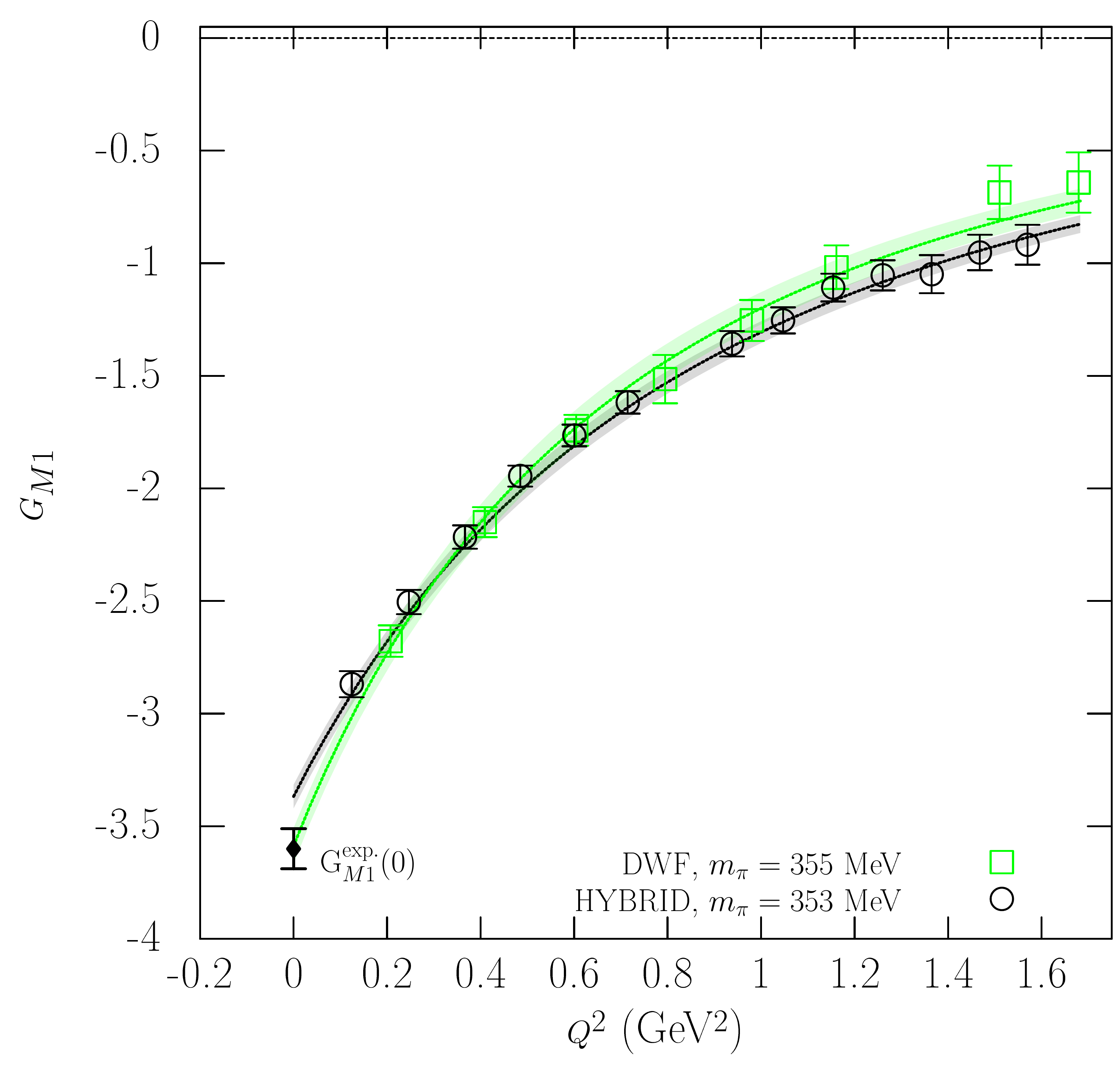}
\vspace*{-0.5cm}\caption{The magnetic form factor $G_{M1}(Q^2)$ comparing the results from the mixed action approach and the DWF lattice at $m_{\pi}\sim 350$~MeV.}
\label{fig:GM1cutoff}
\end{minipage}\hfill
\begin{minipage}{8cm}
\hspace*{-0.2cm}\includegraphics[width=\linewidth,height=\linewidth]{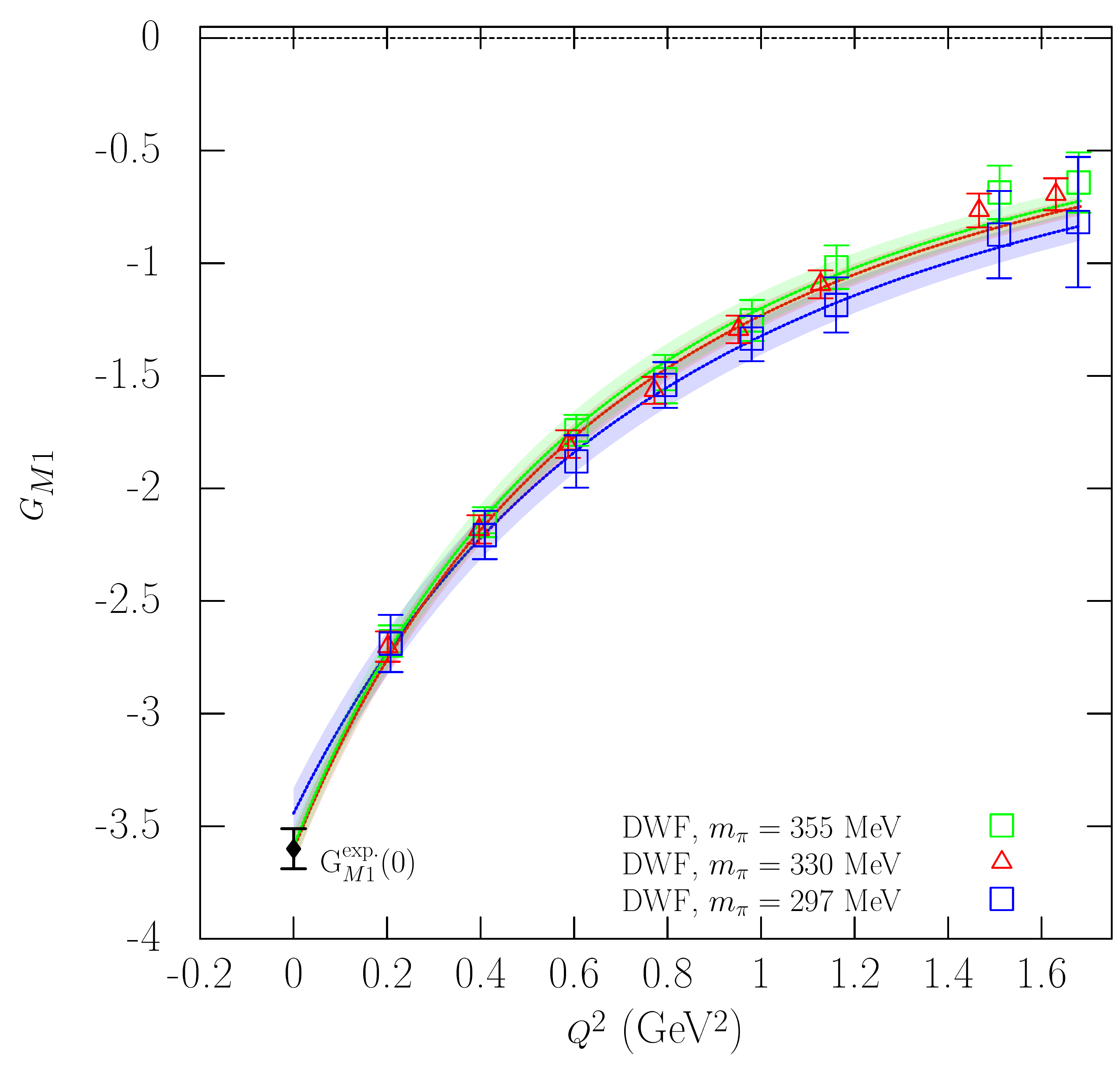}
\vspace*{-0.5cm}
\caption{The magnetic dipole form factor, $G_{M1}$, using DWF  
 at $m_{\pi}=353\ \mathrm{MeV}$,
$m_{\pi}=330\ \mathrm{MeV}$ and $m_{\pi}=297\ \mathrm{MeV}$.
These results are shown  along with the \emph{dipole fit} as given in  Eq.~(\protect \ref{gm1dipoleform}). The  datum for the 
magnetic dipole form factor at 
$Q^2=0\ \mathrm{GeV}^2$, $G^{\mathrm{exp.}}_{M1}(0)=-3.60(8)$, is also included.
	    }\label{fig:GM1}
\end{minipage}
\end{figure}

In order to check for cut-off effects we perform a comparison between the hybrid 
results and results obtained at the same pion mass using DWF on our fine lattice.
This comparison is shown in Fig.~\ref{fig:GM1cutoff}, 
The results using a hybrid action show a small deviation having a smaller
slope as compared to the DWF results. This is the same behavior as was
observed 
 in the case of the electric form factor. 
Given the fact that the lattice spacing for the mixed action is the largest 
this points to cut-off effects. In  Fig.~\ref{fig:GM1}
we show results obtained using DWF on the coarse and fine lattices,
which are in agreement. This  indicates that for these lattice spacings 
 cut-off effects are small. 

\begin{figure}[h]
\hspace{-0.2cm}\includegraphics[width=0.6\linewidth,height=0.5\linewidth]{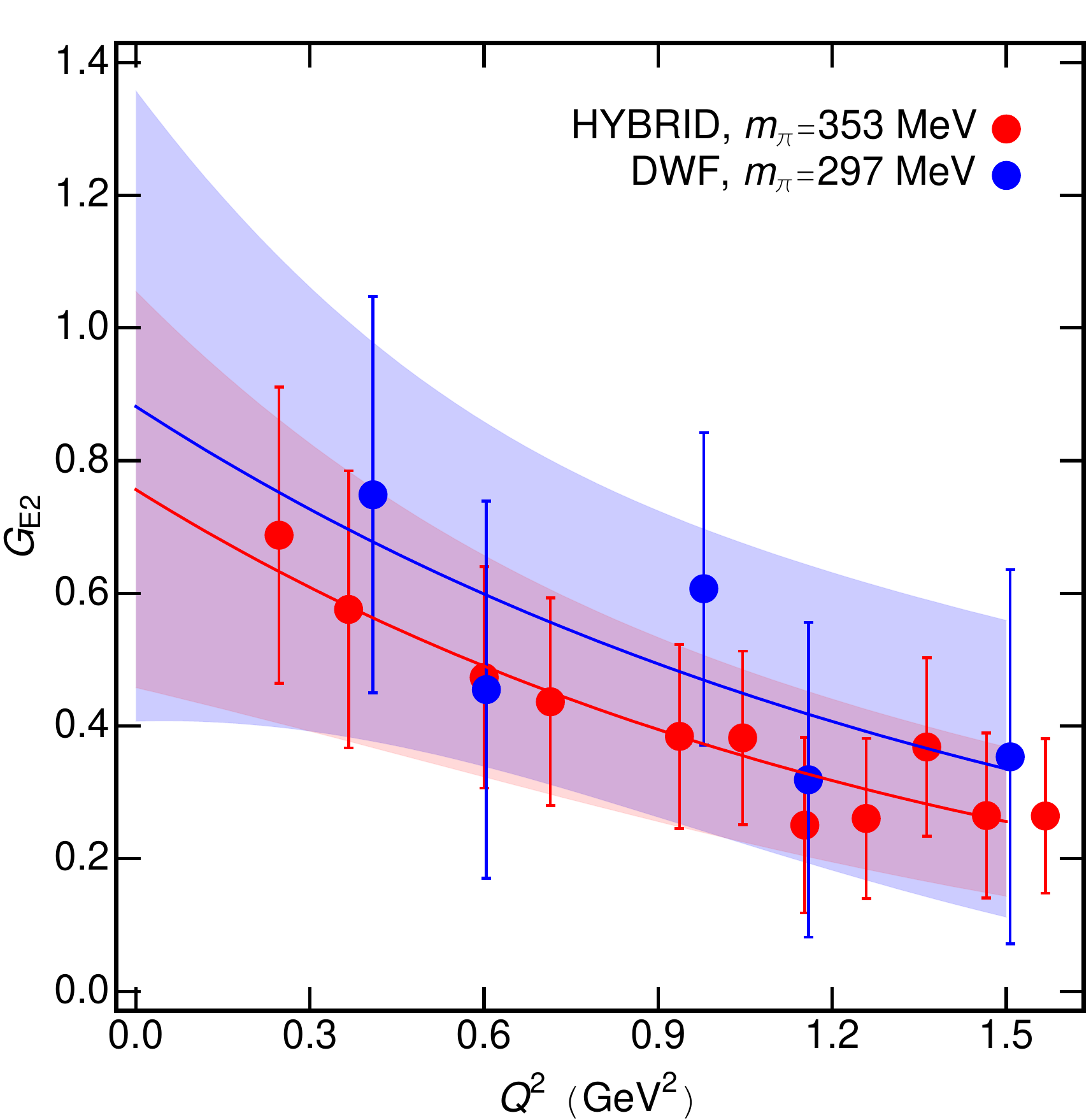}
\vspace*{-0.25cm}\caption{The subdominant electric quadrupole form factor  $G_{E2}(Q^2)$ for DWF using the fine lattice at
$m_{\pi}=297\ \mathrm{MeV}$, and using the
hybrid action at $m_{\pi}=353\ \mathrm{MeV}.$ The extrapolated values at $Q^2=0$ are also depicted. The two results, apart from being consistent within errors indicate a non-zero deformation for the $\Omega^-$ baryon. 
}
\label{figGE2type4}
\end{figure}
The $Q^2$-dependence of the form factors, as in the case of $G_{E0}$, can be described by a dipole form
as can be seen in Figs.~\ref{fig:GM1cutoff} and \ref{fig:GM1}.
 Fitting to the two-parameter  
exponential, dipole and tripole  forms
\begin{align}
\label{gm1expform}
   G_{M1}(Q^2) &= G_0\exp\biggl(-\frac{Q^2}{\Lambda_{M1}^{2}}\biggr),\\
\label{gm1dipoleform}
   G_{M1}(Q^2) &= \frac{G_0}{\big(1+ \frac{Q^2}{\Lambda_{M1}^{2}}\big)^{2}}, \\
\label{gm1tripoleform}
   G_{M1}(Q^2) &= \frac{G_0}{\big(1+ \frac{Q^2}{\Lambda_{M1}^{2}}\big)^{3}},
\end{align}
we can obtain a value for the anomalous magnetic moment of the $\Omega^-$.
\begin{table}[ht]
\begin{center}
\begin{tabular}{cccccc}
\hline\hline 
type of fit &  $\Lambda_{E0}$ [GeV] & ${\chi}_{E0}^2/{\rm d.o.f}$ & $G_0$ &$\Lambda_{M1}$ [GeV] & ${\chi}_{M1}^{2} / {\rm d.o.f}$ \\ 
\hline\hline
\multicolumn{3}{c}{$N_F=2+1$ DWF} ($24^3\times 64$), $N_{\mathrm{confs}}$= 200\\
\hline
exponential & &  & -3.264(89)  & 1.016(27) &   0.225   \\
dipole      & 1.151(13) & 1.500 & -3.601(109) & 1.187(41) &   0.860    \\
tripole     & &  & -3.478(101)  & 1.555(49) &   0.435    \\
\hline
\multicolumn{5}{c}{$N_F=2+1$ DWF} ($32^3\times 64$), $N_{\mathrm{confs}}$=105; $m_{\pi}=355$~MeV\\
\hline
exponential & &  & -3.246(96) & 0.996(43) & 0.159      \\
dipole      &  &  & -3.557(130) & 1.171(63) & 0.440    \\
tripole     & &  & -3.443(116) & 1.530(76) & 0.240    \\
\hline
\multicolumn{5}{c}{$N_F=2+1$ DWF} ($32^3\times 64$), $N_{\mathrm{confs}}$=120; $m_{\pi}=297$~MeV\\
\hline
exponential & &  & -3.199(155) & 1.061(48) & 0.080      \\
dipole      & 1.146(23) & 0.887 & -3.443(173) & 1.277(68) & 0.064    \\
tripole     & &  & -3.355(165) & 1.656(83) & 0.040    \\
\hline
\multicolumn{3}{c}{Hybrid} ($28^3\times 64$), $N_{\mathrm{confs}}$=120 \\
\hline
exponential & & & -3.154(69) & 1.064(30) &  1.147      \\
dipole      & 1.213(17) & 0.168 & -3.368(80) & 1.285(44) &  0.163    \\
tripole     & & & -3.293(76) & 1.662(54) &  0.053    \\
\hline\hline
\end{tabular}
\end{center}
\caption{The fit parameters  for the exponential, dipole and tripole forms
extracted from the lattice data. 
For the fine lattice with $m_\pi=355$ MeV DWF we have only performed inversions for the source type associated with the dominant magnetic dipole form factor $G_{M1}(Q^2)$ (see \protect \Eqref{comb1}). 
}
\label{Table:fitsGM1b}
\end{table}

By utilizing the fit parameter, $G_0 \equiv G_{M1}(0),$ and the lattice computed $\Omega^-$  mass from 
Table~\ref{Table:fitsGM1b}, 
we can evaluate the  magnetic moment
in nuclear magnetons, via the  relation
\begin{align}\label{magmoment}
   \mu_{\Omega^{-}}&=G_0 \bigg(\frac{e}{2m_{\Omega}}\bigg)=G_0\ \bigg(\frac{m_N}{m_{\Omega}}\bigg)\ \mu_N.
\end{align} 
Our value of $\mu_{\Omega^{-}}$ in nuclear magnetons $\mu_N$ is given in
 Table~\ref{ResultsTable}. The values obtained 
are in accord
with  two other recent lattice calculations~\cite{Boinepalli:2009sq,Aubin:2008qp}. 
The calculation in Ref.~\cite{Boinepalli:2009sq} is similar to ours
in the sense that the three-point correlation function is also calculated, but 
the evaluation is
carried out in the quenched theory and only at one value of $Q^2$. 
In Ref.~\cite{Aubin:2008qp} 
a background field method was employed, where
energy shifts were computed using $N_F=2+1$ clover fermions at 
pion mass of 366~MeV on an anisotropic lattice.
\begin{table}[ht]
  \centering
    \begin{tabular}{c c c c c c c c c}
\hline\hline
 & lattice & $m_\pi$ & $Z_V$ & $\mu_{\Omega^{-}}$ &  $\langle r^{2}_{M1}\rangle $ & $\langle r^{2}_{E0}\rangle $  & $G_{E2}(0)$& $Q^{\Omega}_{\frac{3}{2}}$\\
 &{\scriptsize [$L_{s}^{3}\times L_{t}$]}  & {\scriptsize [GeV] }& & {\scriptsize [$\mu_{N}$] } & {\scriptsize  [$\mathrm{fm}^{2}$] } & {\scriptsize  [$\mathrm{fm}^{2}$] } & &{\scriptsize  [$e/m_{\Omega}^{2}$] }\\
\hline
%
This work & HYB: $28^{3}\times 64$& 0.353  & 1.121(2)  & -1.775(52) &0.283(20)    & 0.338(9) & 0.838(19) & -1.366(222)\\
 & DWF: $24^{3}\times 64$         & 0.330  & 0.727(1)   & -1.904(71) & 0.332(23)  & 0.353(8) & --& --\\
 & DWF: $32^{3}\times 64$         & 0.355  & 0.7479(22)  & -1.868(78) &0.341(37) & -- &-- & --\\
 & DWF: $32^{3}\times 64$         & 0.297  & 0.7543(4)  & -1.835(94) &0.286(31) & 0.355(14) & 0.959(41)& -1.892(204)\\
 & extrapolated                  &  0.140 &  --         & -1.875(399) & 0.321(16) & 0.348(52) &0.898(60) & -1.651(262)
\\
\hline
Ref.~\cite{Boinepalli:2009sq} & $20^{3}\times 40$  & 0.697 & 1  & -1.697(65) &-- & 0.307(15) & -- & -- \\
Ref.~\cite{Aubin:2008qp} & $24^{3}\times 128$  & 0.366 & -- & -1.93(8) &-- &-- &--  & --\\
%
\hline
Ref.~\cite{PDG:2008} & --  & -- & -- & -2.02(5) &--   \\
%
\hline\hline
\end{tabular}
\caption{The magnetic moment $\mu_{\Omega^{-}}$, the electric charge and 
magnetic  radii and
 the electric quadrupole moment $Q^{\Omega}_{\frac{3}{2}}$ as
extracted using Eq.~(\ref{eq:quadrup}). The values of 
$\mu_{\Omega^{-}}$,  $\langle r^{2}_{M1}\rangle $, $\langle r^{2}_{E0}\rangle $ and $Q^{\Omega}_{\frac{3}{2}}$ shown above arise from the dipole fit
form. Note that $\langle r^{2}_{M1}\rangle =-\frac{6}{G_{M1}(0)}\frac{dG_{M1}(Q^2)}{dQ^2}\big|_{Q^2=0}. $
} \label{ResultsTable}
\end{table}
\subsection{Electric quadrupole form factor}
From the perspective  of hadron structure,
the extraction of the electric quadrupole form factor is of special interest 
since it can be used to provide valuable information regarding the deformation of a hadron. 
In this work we 
extract for the first time in unquenched QCD  the subdominant $G_{E2}$ form factor for the $\Omega^-$ baryon, to sufficient accuracy to exclude zero values. 
This has been achieved by utilizing two different lattices:
 namely, the fine DWF lattice and the MILC lattice at lattice spacings 
of $a=0.084$~fm and $a=0.124$~fm respectively.
We note that for the coarse DWF lattice the results for $G_{E2}$ 
are too noisy
to exclude a zero value and we
therefore do not present them here.
The lattice results for  $G_{E2}$ 
are depicted in Fig.~\ref{figGE2type4}. 
The value of the quadrupole electric form factor $G_{E2}(Q^2)$ at $Q^2=0$ using the
exponential form to fit the lattice results is 
$0.756(298)$ for the hybrid action and $0.882(475) $ for the fine DWF lattice.
From these results it is readily deduced  that the shape of the
$\Omega^-$ hyperon must deviate from the spherical one. 

The electric quadrupole moment determined from the fits
 as $Q_\Omega=G_{E2}(0)\frac{e}{m_\Omega^2}$ can be related
to the transverse charge density in the infinite momentum frame.
 For instance, the 
transverse charge density defined in the light-front for spin projection 3/2, is
given by~\cite{Alexandrou:2009hs,Alexandrou:2009nj}
\begin{eqnarray}
Q^\Omega_{\frac{3}{2}} 
&=& \frac{1}{2} 
\, \biggl\{ 2 \left[ G_{M1}(0) - 3 e_\Omega \right] + 
\left[ G_{E2}(0) + 3 e_\Omega \right] \biggr\} 
\, \left( \frac{e}{m_\Omega^2} \right). 
\label{eq:quadrup}
\end{eqnarray}

\begin{figure}[h]
\begin{minipage}{7cm}
\includegraphics[width=\linewidth,height=\linewidth]{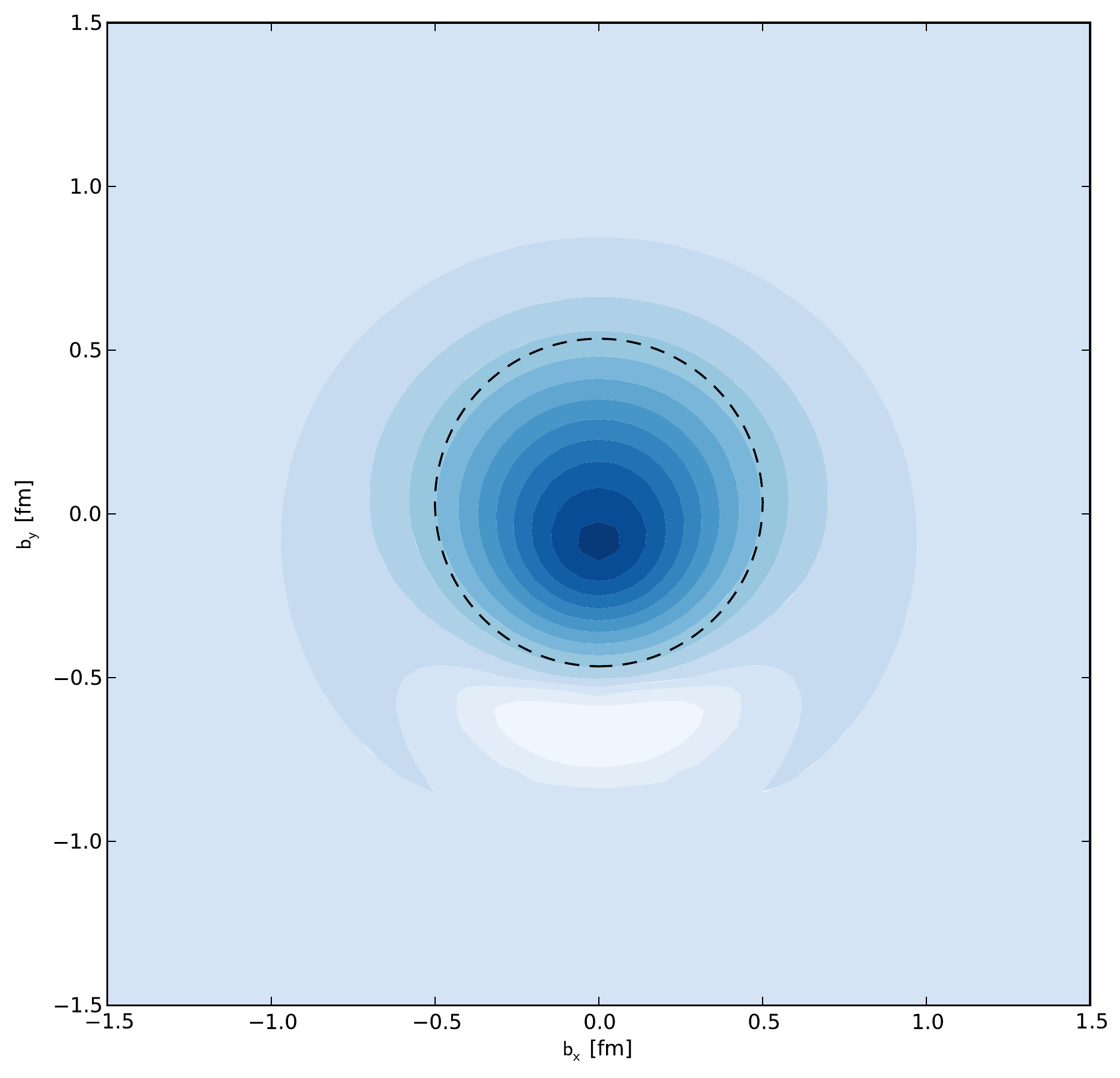}
\end{minipage}\hfill
\begin{minipage}{7cm}
\includegraphics[width=\linewidth,height=\linewidth]{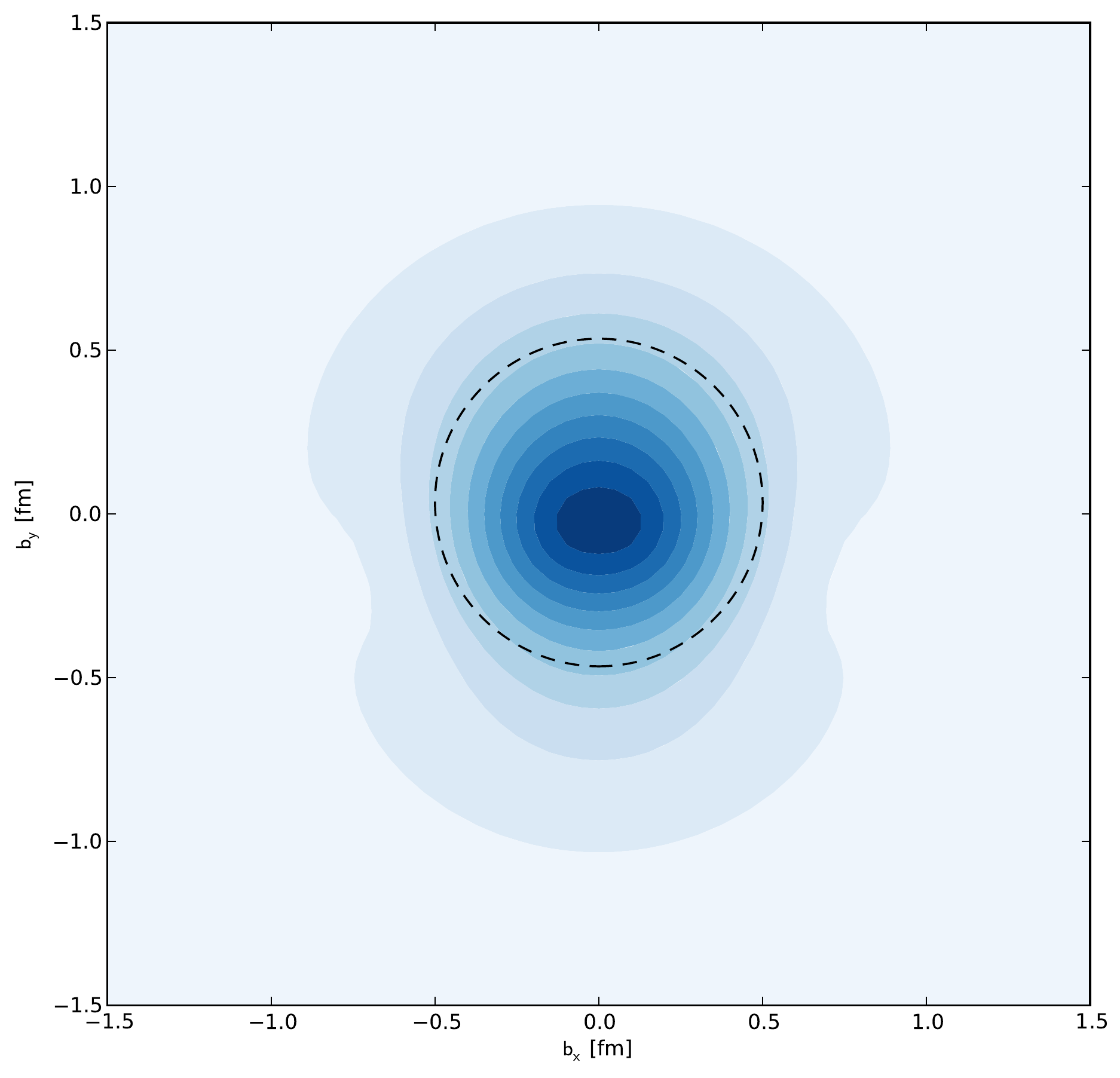}
\end{minipage}\hfill
\vspace*{-0.25cm}\caption{Transverse charge densities in the $\Omega^-$
with polarization along the x-axis. Left: $\rho^\Omega_{T 3/2}(\vec{b})$. 
Right:$\rho^\Omega_{T 1/2}(\vec{b})$. A circle of radius 0.5 fm is drawn in
order to clearly demonstrate the deformation. For the evaluation of
the densities we used the dipole parametrization of the form factors.}
\label{fig:transverse_den}
\end{figure}

\begin{figure}[ht]
\begin{minipage}{8cm}
\includegraphics[width=0.6\linewidth,height=0.6\linewidth]{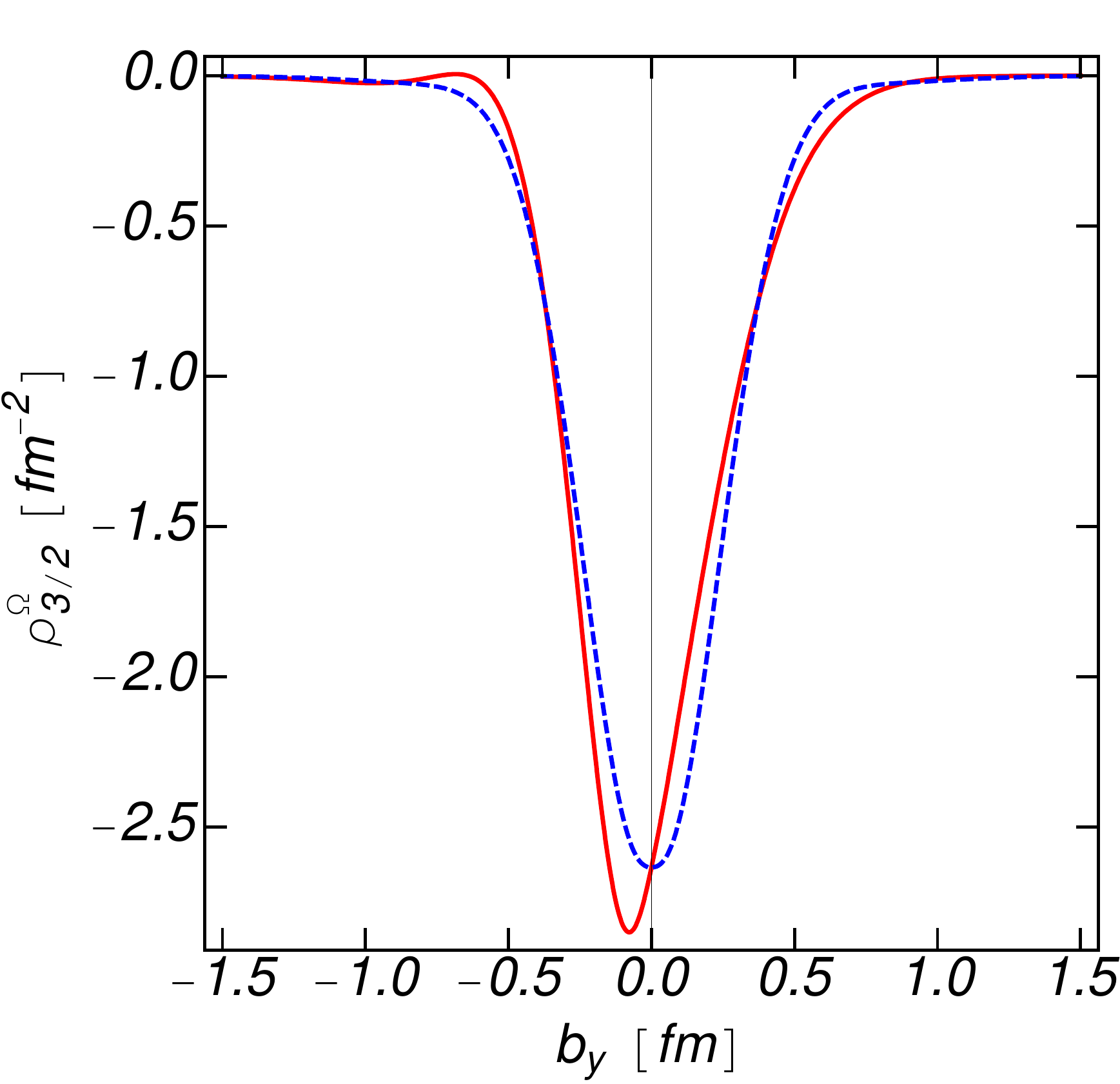}
\end{minipage}\hfill
\begin{minipage}{8cm}
\includegraphics[width=0.6\linewidth,height=0.6\linewidth]{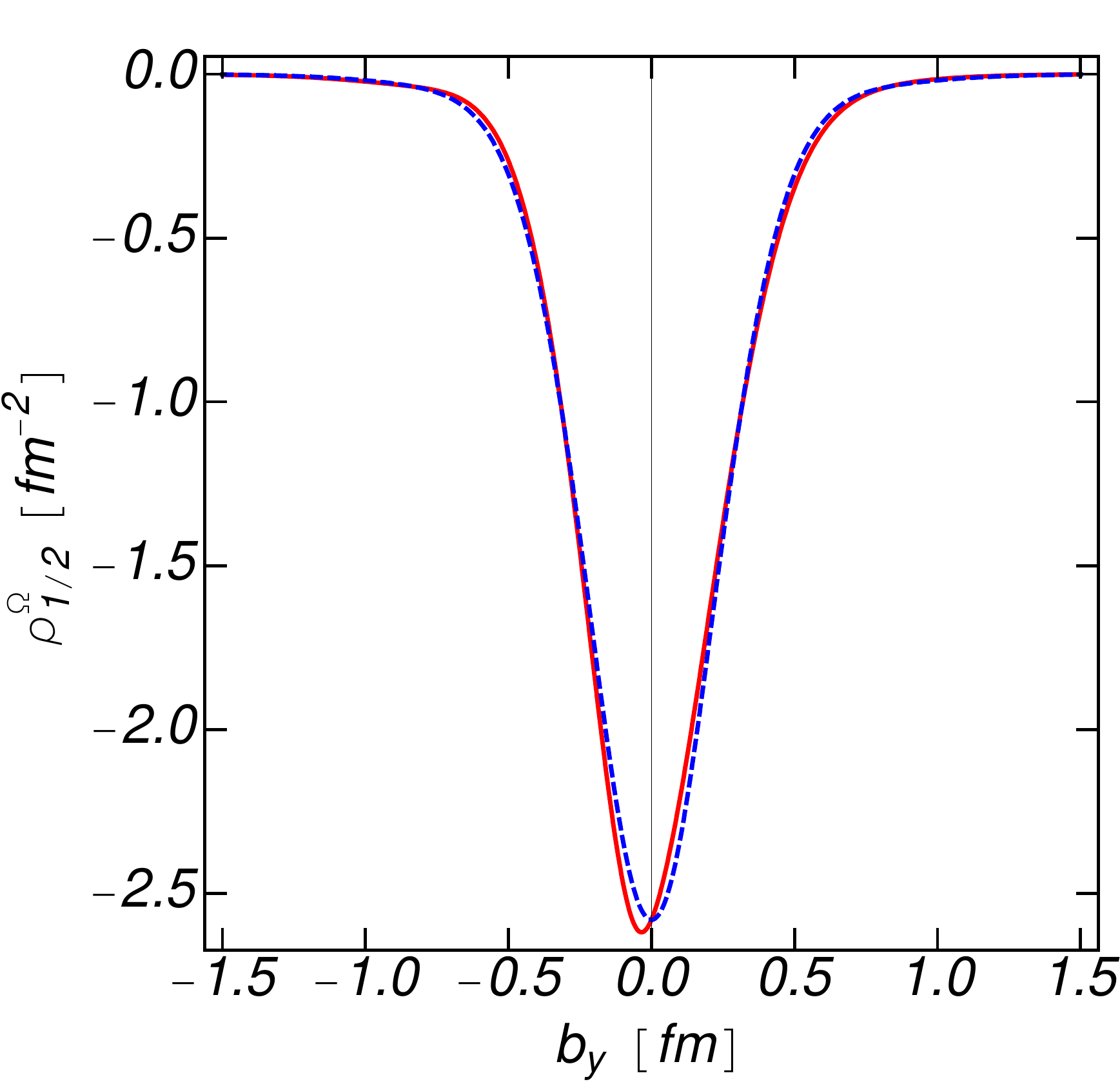}
\end{minipage}\hfill
\vspace*{-0.25cm}\caption{Comparison of the transverse charge densities 
$\rho^\Omega_{T 3/2}(\vec{b})$ (left) and $\rho^\Omega_{T 1/2}(\vec{b})$ (right) along the y-axis to the monopole-field (symmetric) shown by the dashed line.}
\label{fig:tranverse_2Dden}
\end{figure}

We note that for a spin-$\frac{3}{2}$ particle without internal structure, 
for which $G_{M1}(0) = 3 e_\Omega$ and $G_{E2}(0) = -3 e_\Omega$~\cite{Alexandrou:2009hs,Alexandrou:2009nj}, the quadrupole moment of the 
transverse charge densities vanishes. We calculate this quantity 
by using a fit to the electric quadrupole to obtain the value at $Q^2=0$.
The results obtained are shown in Table~\ref{ResultsTable} and plotted in
Fig.~\ref{fig:chiral} for the dipole 
fitting Ansatz. Both of the two values are negative and consistent within statistical errors. Therefore, they suggest that the quark charge distribution in the $\Omega^-$ must be deformed. In order to investigate the deformation in
more detail we construct the transverse charge density in the infinite momentum
frame, following Refs.~\cite{Alexandrou:2009hs,Alexandrou:2009nj}. Considering
the spin of the $\Omega$ along the x-axis and 
states of transverse spin $s_\perp=3/2$ and $s_\perp=1/2$ we obtain 
the transverse charge densities $\rho^\Omega_{T 3/2}(\vec{b})$ and  $\rho^\Omega_{T 1/2}(\vec{b})$ in term of the two-dimensional impact parameter $\vec{b}$.
 In Fig.~\ref{fig:transverse_den} we compare   $\rho^\Omega_{T 3/2}(\vec{b})$ and  $\rho^\Omega_{T 1/2}(\vec{b})$. As can be seen, in a state 
of transverse spin projection $s_\perp=3/2$ the $\Omega^-$ shows a
small elongation along the spin axis (prolate)~\footnote{Note that this
is consistent with the negative sign of $Q_{3/2}$ since the $\Omega^-$ is
negatively charged and have included its charge in the electromagnetic 
current}. This elongation is less
as compared to that seen for the $\Delta^+$. As in the case of the $\Delta^+$, 
in a state
of transverse spin projection $s_\perp=1/2$ the $\Omega^-$ is elongated 
along the axis perpendicular to the spin.

\begin{figure}[ht]
\includegraphics[width=0.7\linewidth,height=0.7\linewidth]{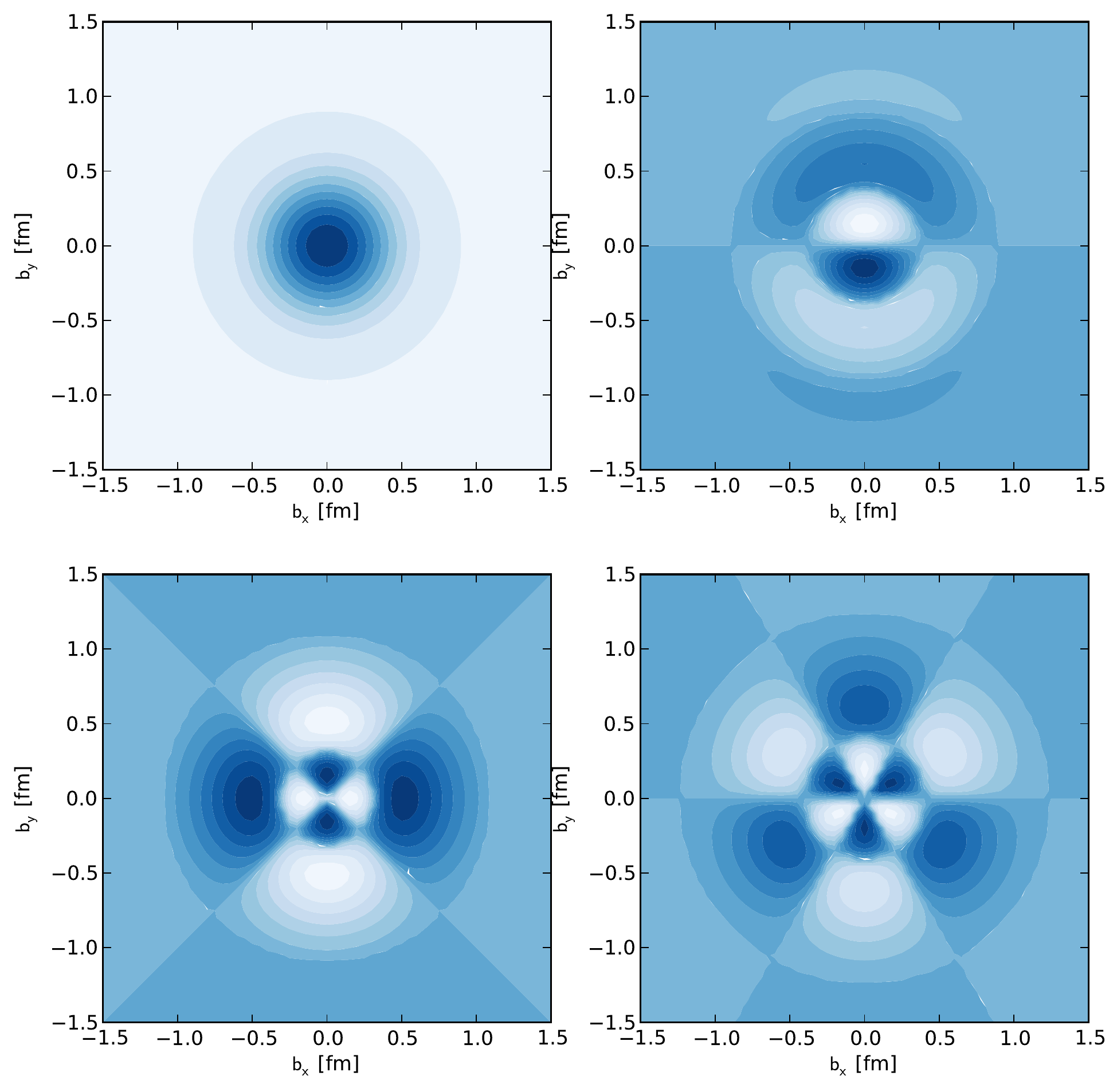}
\vspace*{-0.25cm}\caption{The individual multipoles for the 
transverse charge density  $\rho^\Omega_{T 3/2}(\vec{b})$ in the $\Omega^-$
with polarization along the x-axis.Upper left: monopole-field. Upper right: dipole-field. Lower left: quadrupole-field. Lower right: Octupole-field}
\label{fig:multipoles3ov2}
\end{figure}
In Fig.~\ref{fig:tranverse_2Dden} we show the profile of the transverse densities compared to the monopole field that is symmetric. 
In Fig.~\ref{fig:multipoles3ov2} we show the individual multipole fields
for the state with transverse spin $s_\perp=3/2$.

\subsection{Extrapolation to the physical point}
\begin{figure}[ht]
\begin{center}
\hspace{-0.2cm}\includegraphics[width=0.5\linewidth,height=0.65\linewidth]{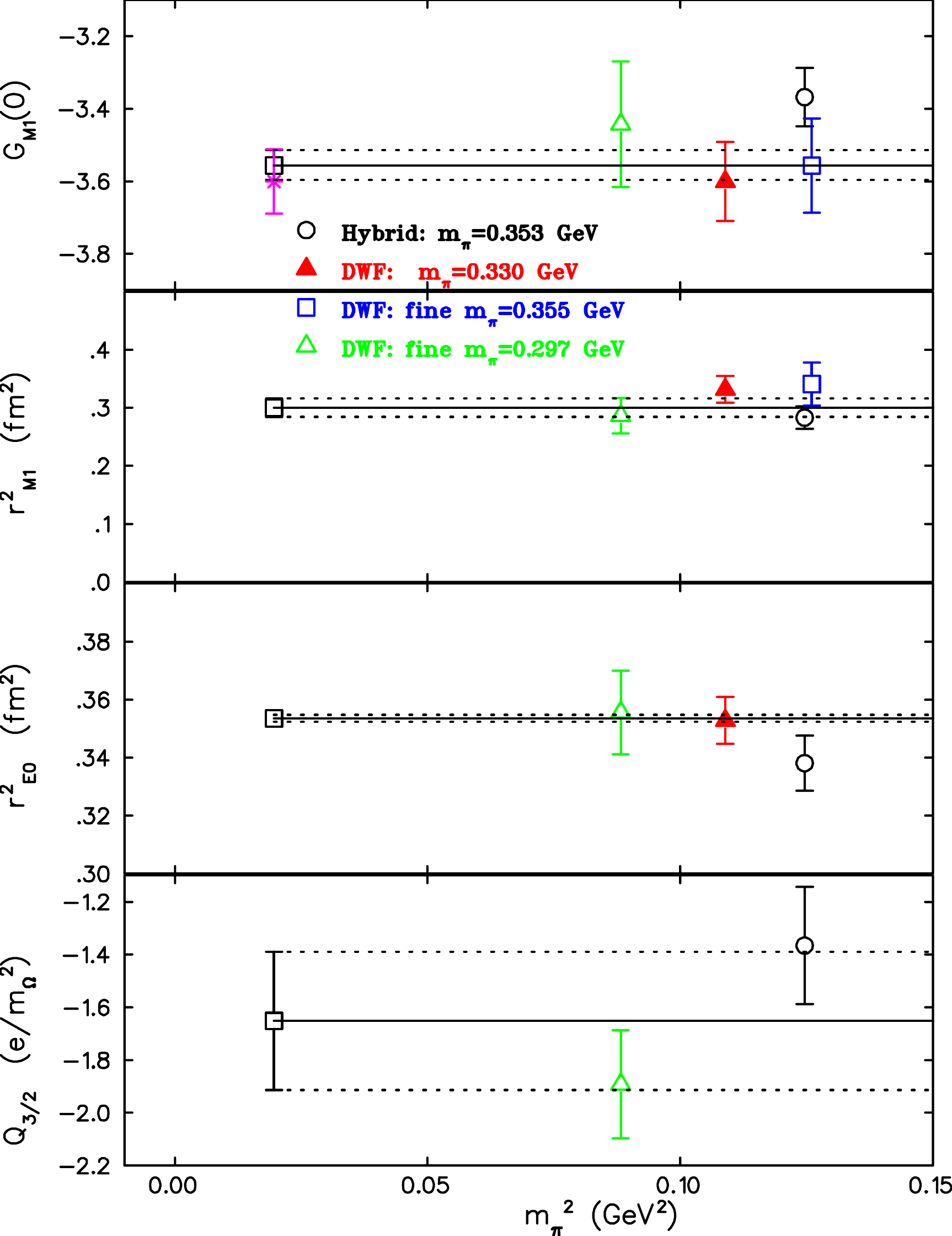}
\vspace*{-0.4cm}\caption{
From top to bottom we show $G_{M1}(0)$,
the magnetic radius $\langle r^2_{M1} \rangle$,
the electric radius $\langle r^2_{E0} \rangle$ and
the quadrupole moment
extracted from Eq.~(\ref{eq:quadrup}) as a function of $m_\pi^2$
 extracted from  dipole fits. 
The point shown by the filled square is the value extracted
from the fit at the physical pion mass. In all cases except for the quadrupole moment
the results using the hybrid action are excluded form the fit.
}
\label{fig:chiral}
\end{center}
\end{figure}

In this section we examine the sea quark dependence of the magnetic moment,  radii and the quadrupole moment.
They are extracted by fitting the $Q^2$-dependence 
of the form factors to a dipole form. As can be seen from Fig.~\ref{fig:chiral} the sea quark mass dependence is consistent with a constant for all quantities
confirming that sea quark effects are small.
In particular, the value of the magnetic form factor at $Q^2=0$ is consistent with experiment. On the other hand extrapolating
the magnetic moment we obtain the value given in Table~\ref{ResultsTable}. This is 5\% smaller than experiment which
is to be expected given the larger value of the strange quark mass. 
The reason is that the mass of the $\Omega^-$ is 5\% larger than experiment  and this will affect
the value of the magnetic moment when we convert to nuclear magnetons.
 In the fits for the magnetic moment and radii we did not include the results obtained in the hybrid action 
because of the small finite $a$ effects observed. Given the large statistical errors
on  quadrupole moment such small finite-$a$ effects are negligible and therefore, in this case, we 
include the result using the hybrid action to obtain the extract the value at the physical point. 
 In Table~\ref{ResultsTable} we give the values that we find
at the physical point for the radii and the dipole and quadrupole moments of the 
transverse charge density obtained from Eq.~(\ref{eq:quadrup}).
\section{Conclusions}
By utilizing properly constructed sequential sources the dominant $\Omega^-$ electromagnetic form factors $G_{E0}$ and $G_{M1}$ are calculated  with good accuracy using dynamical domain-wall fermion configurations as well as
a hybrid action. 

In addition, we   extract the magnetic moment of the $\Omega^-$ 
by fitting the magnetic dipole form factor $G_{M1}$
 to a two-parameter dipole form. We find a value that is within errors  
to the  experimentally measured value~\cite{PDG:2008}.
The electric charge and magnetic radii
($\langle r^{2}_{E0}\rangle $ and $\langle r^{2}_{M1}\rangle$)
are computed and like the magnetic dipole moment they do not show
sea quark dependence  in the range of masses studied in this work.

Finally, the subdominant electric quadrupole 
form factor $G_{E2}$ is computed for the first time
in an unquenched
 lattice calculation to sufficient accuracy to exclude a zero value. 
This has been accomplished  by constructing an appropriate sink
that isolates it from the two dominant form factors. We find consistent results
with DWF and using a hybrid action.
The positive non-zero values of $G_{E2}$ at $Q^2=0$ 
suggest that the structure of the $\Omega^-$ baryon is non-spherical. 
In the light-front frame we find that the quark charge density in a $\Omega^-$ for a state of 
transverse spin projection +3/2 is shows an elongation along the axis of the 
spin (prolate deformation).  As compared to the $\Delta^+$ in the same state
the amount of deformation seen in the $\Omega^-$ is smaller.
\newline
\centerline{\bf Acknowledgments}

 This research was partly supported by the Cyprus Research Promotion Foundation (R.P.F) under contracts $\mathrm{\Pi}$ENEK/ENI$\mathrm{\Sigma}$X/0505-39 and EPYAN/0506/08 and by the U.S. Department of Energy under Grant  DE-FG02-94ER-40818. The authors would also like to acknowledge the use of dynamical domain wall fermions configurations provided by the RBC-UKQCD collaborations and the use of Chroma software~\cite{Edwards:2004sx}.
Part of the computational resources required for these calculations where 
provided by the J\"ulich Supercomputing Center at Research Center J\"ulich.


\begin{thebibliography}{99}
 \bibitem{PDG:2008}
  C.~Amsler \emph{et al}. (Particle Data Group), PL B\textbf{667} (2008) 1.

\bibitem{Alexandrou:2008bn}
  C.~Alexandrou {\it et al.},
  Phys.\ Rev.\  D {\bf 79} (2009) 014507.

\bibitem{Orginos:1999cr}
  K.~Orginos, D.~Toussaint and R.~L.~Sugar  [MILC Collaboration],
  Phys.\ Rev.\  D {\bf 60}, 054503 (1999)
  [arXiv:hep-lat/9903032].

\bibitem{Nozawa:1990gt}
  S.~Nozawa and D.~B.~Leinweber,
  Phys.\ Rev.\  D {\bf 42} (1990) 3567.

\bibitem{Allton:2008pn}
  C.~Allton {\it et al.}  [RBC-UKQCD Collaboration],
  Phys.\ Rev.\  D {\bf 78} (2008) 114509.

\bibitem{Syritsyn:2009mx}
  S.~N.~Syritsyn {\it et al.},
  Phys.\ Rev.\  D {\bf 81}, 034507 (2010)
  [arXiv:0907.4194 [hep-lat]].

\bibitem{Bernard:2001av}
  C.~W.~Bernard {\it et al.},
  Phys.\ Rev.\  D {\bf 64}, 054506 (2001)
  [arXiv:hep-lat/0104002].

\bibitem{Aubin:2004ck}
  C.~Aubin {\it et al.}  [HPQCD Collaboration and MILC Collaboration and
                  UKQCD Collaboration],
  Phys.\ Rev.\  D {\bf 70}, 031504 (2004)
  [arXiv:hep-lat/0405022].

\bibitem{WalkerLoud:2008bp}
  A.~Walker-Loud {\it et al.},
  Phys.\ Rev.\  D {\bf 79}, 054502 (2009)
  [arXiv:0806.4549 [hep-lat]].

\bibitem{Renner:2004ck}
  D.~B.~Renner {\it et al.}  [LHP Collaboration],
  Nucl.\ Phys.\ Proc.\ Suppl.\  {\bf 140}, 255 (2005)
  [arXiv:hep-lat/0409130].

\bibitem{Hagler:2007xi}
  Ph.~Hagler {\it et al.}  [LHPC Collaborations],
  Phys.\ Rev.\  D {\bf 77}, 094502 (2008)
  [arXiv:0705.4295 [hep-lat]].

\bibitem{Aoki:2010xg}
  Y.~Aoki {\it et al.},
  arXiv:1003.3387 [hep-lat].

\bibitem{Alexandrou:1992ti}
  C.~Alexandrou, S.~Gusken, F.~Jegerlehner, K.~Schilling and R.~Sommer,
  Nucl.\ Phys.\  B {\bf 414} (1994) 815.

\bibitem{APEsmearing}
M. Albanese \emph{et. al.} (APE Collaboration) 
Phys. \ Lett. \ B {\bf 192} (1987) 163.

\bibitem{Montvay:1994cy}
  I.~Montvay and G.~Munster,
{\it  Cambridge, UK: Univ. Pr. (1994) 491 p. (Cambridge monographs on mathematical physics)}

\bibitem{Dolgov:2002zm}
  D.~Dolgov {\it et al.}  [LHPC collaboration and TXL Collaboration],
  Phys.\ Rev.\  D {\bf 66}, 034506 (2002)
  [arXiv:hep-lat/0201021].


\bibitem{Aoki:2007xm}
  Y.~Aoki {\it et al.},
  Phys.\ Rev.\  D {\bf 78} (2008) 054510.

\bibitem{Bratt:2010jn}
  J.~D.~Bratt {\it et al.}  [LHPC Collaboration],
  arXiv:1001.3620 [hep-lat].

\bibitem{Boinepalli:2009sq}
  S.~Boinepalli, D.~B.~Leinweber, P.~J.~Moran, A.~G.~Williams, J.~M.~Zanotti and J.~B.~Zhang,
 Phys. Rev. D 80, 054505 (2009), arXiv:0902.4046 [hep-lat].

\bibitem{Aubin:2008qp}
  C.~Aubin, K.~Orginos, V.~Pascalutsa and M.~Vanderhaeghen,
 Phys. Rev. D 79, 051502 (2009), arXiv:0811.2440 [hep-lat].

\bibitem{Alexandrou:2009hs}
  C.~Alexandrou {\it et al.},
  Nucl.\ Phys.\  A {\bf 825} (2009) 115.

\bibitem{Alexandrou:2009nj}
  C.~Alexandrou {\it et al.},
  PoS C {\bf D09}, 092 (2009)
  [arXiv:0910.3315 [hep-lat]].

\bibitem{Edwards:2004sx}
  R.~G.~Edwards and B.~Joo  [SciDAC Collaboration and LHPC Collaboration and
                  UKQCD Collaboration],
  Nucl.\ Phys.\ Proc.\ Suppl.\  {\bf 140}, 832 (2005)
  [arXiv:hep-lat/0409003].



\end{thebibliography}
\end{document}